\documentclass{aa}
\usepackage[varg]{txfonts}
\usepackage{graphicx}
\usepackage{txfonts}
\usepackage[]{hyperref}
\usepackage{natbib}
\usepackage{color}
\def\simle{\mathrel{\hbox{\rlap{\hbox{\lower4pt\hbox{$\sim$}}}\hbox{$<$}}}}
\def\simgr{\mathrel{\hbox{\rlap{\hbox{\lower4pt\hbox{$\sim$}}}\hbox{$>$}}}}

\titlerunning{
An absence of binary companions to SMC WR stars: implications for mass loss and $M_\mathrm{BH}$ at low $Z$
}

\begin{document}

\title{
An absence of binary companions to Wolf-Rayet stars in the Small Magellanic Cloud
\thanks{Based on observations collected at the European Organisation for Astronomical Research in the Southern Hemisphere under ESO programme 108.22M1.001. 
}}
\subtitle{
Implications for mass loss and black hole masses at low metallicity
}

\author{A. Schootemeijer\inst{1} \and T. Shenar\inst{2} \and N. Langer\inst{1,3} \and N. Grin\inst{1} \and H. Sana\inst{4} \and G. Gr\"{a}fener\inst{1} \and C. Sch\"{u}rmann\inst{1} \and C. Wang\inst{5} \and X.-T. Xu\inst{1}
}

\institute{Argelander-Institut f\"{u}r Astronomie, Universit\"{a}t Bonn, Auf dem H\"{u}gel 71, 53121 Bonn, Germany\\  \email{aschoot@astro.uni-bonn.de} \and
The School of Physics and Astronomy, Tel Aviv University, Tel Aviv 6997801, Israel \and
Max-Planck-Institut für Radioastronomie, Auf dem Hügel 69, 53121 Bonn, Germany \and 
Institute of Astronomy, KU Leuven, Celestijnenlaan 200D, 3001 Leuven, Belgium \and Max Planck Institute for Astrophysics, Karl-Schwarzschild-Strasse 1, 85748 Garching, Germany}

\abstract{
{In order to predict the black hole mass distributions at high redshift, we need to understand whether very massive single stars ($M \simgr 40$\,M$_\odot$) at low metallicity $Z$ lose their hydrogen-rich envelopes, like their metal-rich counterparts, or whether a binary companion is required to achieve this.
}
{To test this, we undertake a deep spectroscopic search for binary
companions of the seven apparently single Wolf-Rayet (WR) stars in the Small Magellanic Cloud (SMC; where $Z \simeq 1/5\,Z_\odot$). 
}
{For each of them, we acquired six high-quality VLT-UVES spectra spread over a time period of 1.5 years. By using the narrow N\,{\sc v} lines in these spectra, we monitor radial velocity (RV) variations to search for binary motion.}
{We find low RV variations between 6 and 23\,km/s for the seven WR stars, with a median standard deviation of $5$\,km/s. 
Our Monte Carlo simulations imply probabilities below $\sim$5\% for any of our target WR stars to have a binary companion more massive than $\sim$5\,M$_\odot$ at orbital periods of less than a year.
We estimate that the probability that \textit{all} our target WR stars have companions with orbital periods shorter than 10\,yr is below $\sim$10$^{-5}$,
and argue that the observed modest RV variations may originate from intrinsic atmosphere or wind variability.
}
{Our findings imply that metal-poor massive stars born with $M \gtrsim 40$\,M$_\odot$ can lose most of their hydrogen-rich envelopes via stellar winds or eruptive mass loss, which strongly constrains their
initial mass -- black hole mass relation.
We also identify two of our seven target stars (SMC\,AB1 and SMC\,AB11) as runaway stars with a peculiar radial velocity of $\sim$80\,km/s.
Moreover, with all five previously detected WR binaries in the SMC exhibiting orbital periods of below 20\,d, a puzzling absence of intermediate-to-long-period WR binaries has emerged, with strong implications for the outcome of massive binary interaction at low metallicity.
}
}


\maketitle

\section{Introduction} \label{sec:intro}

Two recent developments in observational astronomy have sparked a great interest in understanding very massive stars ($M>40$\,M$_\odot$) in the Early Universe. First, the gravitational wave (GW) observatories of the LIGO/Virgo collaboration have detected black hole (BH) mergers of tens of Solar masses \citep{Abbott16}. These BHs were most likely formed at low metallicity \citep{Giacobbo18}. Second, the James Webb Space Telescope just started a new era in observational astronomy, and has already delivered spectacular discoveries of bright blue galaxies born briefly after the Big Bang \citep{Castellano22, Adams23}.

Very massive stars are crucial for both of the above topics, as they are progenitors of BH-BH mergers and drivers of galaxy evolution. One question is of particular relevance here: could very massive stars in the early Universe lose their hydrogen-rich envelopes via stellar winds? Current stellar evolution models give discrepant results \citep{Yusof13, Szecsi15, Pauli22}. If winds were strong enough for this, the consequences are, i), a limitation of the mass of BHs produced by isolated stars, and, ii), the formation of isolated hot Wolf-Rayet (WR) stars, which may chemically enrich and ionize their environment. If not, massive stars in the early universe without a binary companion remained massive and left behind massive black holes. Indeed, for stars born more massive than 30\,M$_\odot$, wind mass loss is the major uncertainty regarding their final mass as BHs \citep{Fryer12}.
Also, massive stars in the early universe would then probably needed companions to become WR stars.

A one-of-a-kind environment to study massive stars with chemical compositions similar to those in the Early Universe
is the Small Magellanic Cloud (SMC), where the metallicitity ($Z$) is about five times lower than Solar \citep{Venn99}. 
As a satellite galaxy of the MW only 62\,kpc away \citep{Graczyk20} it is close enough to study individual stars.
There are twelve WR stars in the SMC. Five of them are known to have a hot and massive companion, and for the seven others previous searches did not find any companion stars \citep{Westerlund64, Smith68, Sanduleak68, Sanduleak69, Breysacher78, Azzopardi79, Moffat82, Moffat85, Moffat88, morgan91, Bartzakos01, Massey01, Massey03, Foellmi03, Foellmi04, Hainich15, Shenar16, Shenar18, Neugent18}. 


\begin{figure}[t]
\begin{center}
\includegraphics[width=\linewidth]{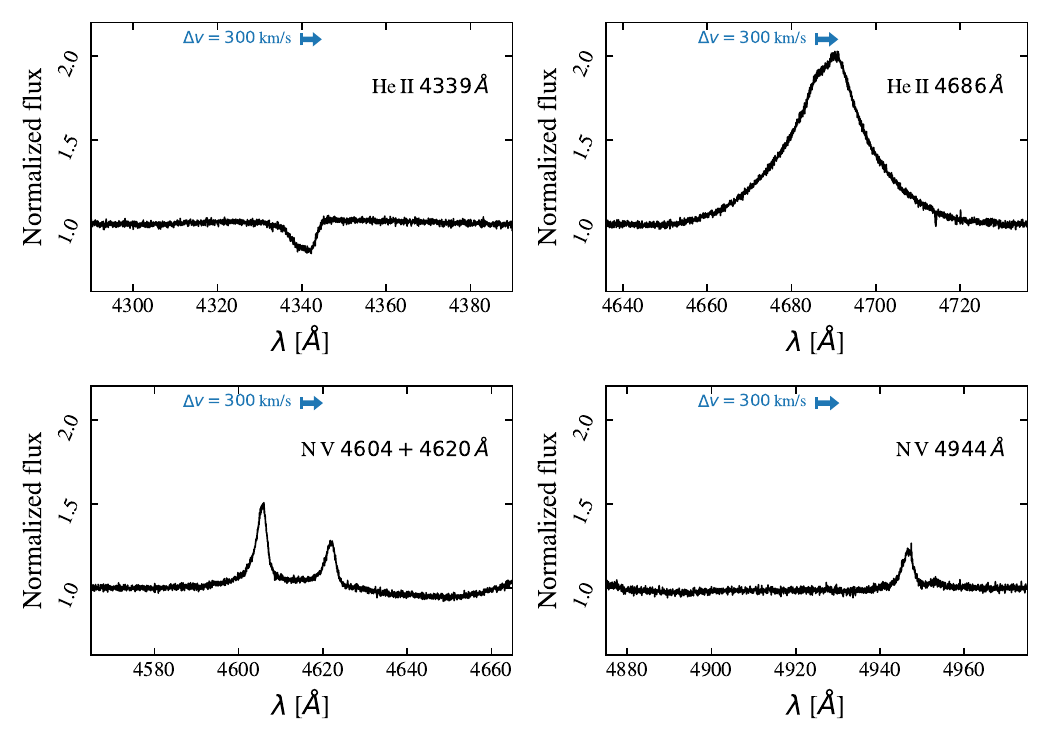} 
 \caption{Normalized 100\,\AA\ wide cutouts of the co-added VLT-UVES spectrum of SMC\,AB9. For reference, the blue arrow indicates the wavelength shift resulting from a 300\,km/s velocity shift at the central wavelength of each panel.}
 \label{fig:ab9_lines}
\end{center}
\end{figure}

To search for binary companions, \cite{Foellmi03} -- hereafter F03 -- have monitored the radial velocity (RV) of all SMC WR stars except SMC\,AB\,12, which was discovered later \citep{Massey03, Foellmi04} and has not yet been subjected to RV monitoring, and SMC\,AB8, which is the only WO-type SMC WR star and part of a binary system. F03 presented orbital solutions for the binary stars and investigated the completeness of their study. They showed that for binaries with orbital periods up to 300 days that consist of a 10\,M$_\odot$ WR star with a 30\,M$_\odot$ companion, 
the projected semi-amplitude of the RV of the WR star is typically larger than 30\,km/s, which is the minimum value for which they claim to be able to detect orbital motion. This threshold of 30\,km/s is estimated as $\sqrt{2}\sigma_\mathrm{RV}$, where $\sigma_\mathrm{RV}$ is the measurement error with a reported value of 20\,km/s.

However, this does not rule out the presence of companion stars in general, since different orbital configurations are possible that would decrease 
the projected orbital velocity of the WR star. 
First, WR masses in the SMC are typically at least 20\,M$_\odot$ \citep{Hainich15, Shenar16, Schootemeijer18} instead of the 10\,M$_\odot$ assumed by F03. Second, interaction can take place for wider binaries that have orbital periods up to about 10 years \citep{Sana12}. Third, the companion might be significantly lighter than 30\,M$_\odot$. 

Therefore, with the knowledge gathered up to now it is possible that most or even all SMC WR stars are in binaries, but that at least a few have escaped detection due to long orbital periods and/or relatively low-mass companions \citep{Schootemeijer18}. 
Here, we present a modern RV monitoring survey of the seven apparently-single WR stars in the SMC using data acquired with the 8-meter Very Large Telescope (VLT).  These data allow us to reach conclusive statements regarding the binary status of the apparently single SMC WR stars. This provides important constraints on the ability of isolated massive stars in low-Z environments to shed their outer H-rich layers.

Our paper is organized as follows. In Sect.\,\ref{sec:meth}, we present the VLT data used in this study and our method to measure RVs; the results of these measurements are shown in Sect.\,\ref{sec:results}. We discuss the impact of line profile variability on our results in Sect.\,\ref{sec:lpv}, the sensitivity of our campaign to binary motion in Sect.\,\ref{sec:possible_presence_of_companions}, and other possible signatures of binary companions in Sect.\,\ref{sec:other_signatures}. Finally, we discuss our findings in Sect.\,\ref{sec:discussion} and present our conclusions in Sect.\,\ref{sec:conclusions}.

\section{Methods \label{sec:meth}}
\subsection{Observations \label{sec:obs}}
We made use of high-resolution spectra taken by the Ultraviolet and Visual Echelle Spectrograph \citep[UVES;][]{Dekker00} instrument of the VLT. For our monitoring program (program ID: 108.22M1.001; PI: Schootemeijer) six spectra in total were obtained in Service Mode for each of our seven targets. This was done during three semesters, between October 2021 and December 2022.
We used the standard DIC\,2 437+760 setting and a slit width of 1". The ESO CPL pipeline (version 7.1.2) was used for the reduction of the spectra.
The wavelength calibration was done with the standard ESO procedure that uses a thorium-argon lamp. 
We corrected the spectra for barycentric motion by subtracting the RV measured (as described in Sect\,\ref{sec:rv_measurements}) for the extremely narrow Ca\,{\sc ii} interstellar absorption feature around 3970\,\AA. 
This correction also serves as wavelength calibration.
As a test, we also corrected for barycentric motion using the {\tt baryCorr} function from the \textsc{PyAstronomy} package \textsc{pyasl} \citep{Piskunov02}. Both approaches provided similar results. We chose to use the interstellar Ca\,{\sc ii} method because it yielded RVs that are slightly more constant in time.

The UVES instrument splits the collected light into two arms: a blue arm and a red arm. For the blue arm, the covered wavelength range was 3730--5000\,\AA, the typical signal to noise ratio \citep[S/N; per resolution element, and measured with the method of][]{Stoehr08} of the spectra was 40--50. The blue spectra have a resolving power of $R = \lambda/ \Delta\lambda \approx 41000$, which translates into a bin size of $\Delta \lambda = 0.03$\,\AA\ around 4500\,\AA. 
The lines that we use for the RV measurements are in the wavelength range of the blue arm. The spectra taken with the red arm in the wavelength range 5660--9460\,\AA\ had a typical S/N of 25--30 and a resolving power of $R \approx 42000$. 

SMC\,AB11 has a red source with a similar $G$ magnitude (both have $G=15.8$) 1" away from it, causing the point spread functions of the two objects to blend. To enable a relatively clean extraction of the WR spectrum, the slit orientation for SMC\,AB11 was set to include both sources. The spectrum of the WR star was then extracted from the 2D spectrum image by identifying the spatial pixel beyond which the contribution of the red sources becomes negligible.

For three out of seven targeted stars -- SMC\,AB1, 2, and 4 -- archival UVES spectra from August 2006 \citep[PI: Foellmi; see][]{Marchenko07} 
exist on the ESO archive. In their campaign, a total of 10--12 spectra was taken during two successive nights for each of these three stars. These spectra are in the wavelength range 3930--6030\,\AA, and they have an S/N of $\sim$50 at a resolving power of $R \approx 30000$.

\begin{figure}[t]
\begin{center}
\includegraphics[width=\linewidth]{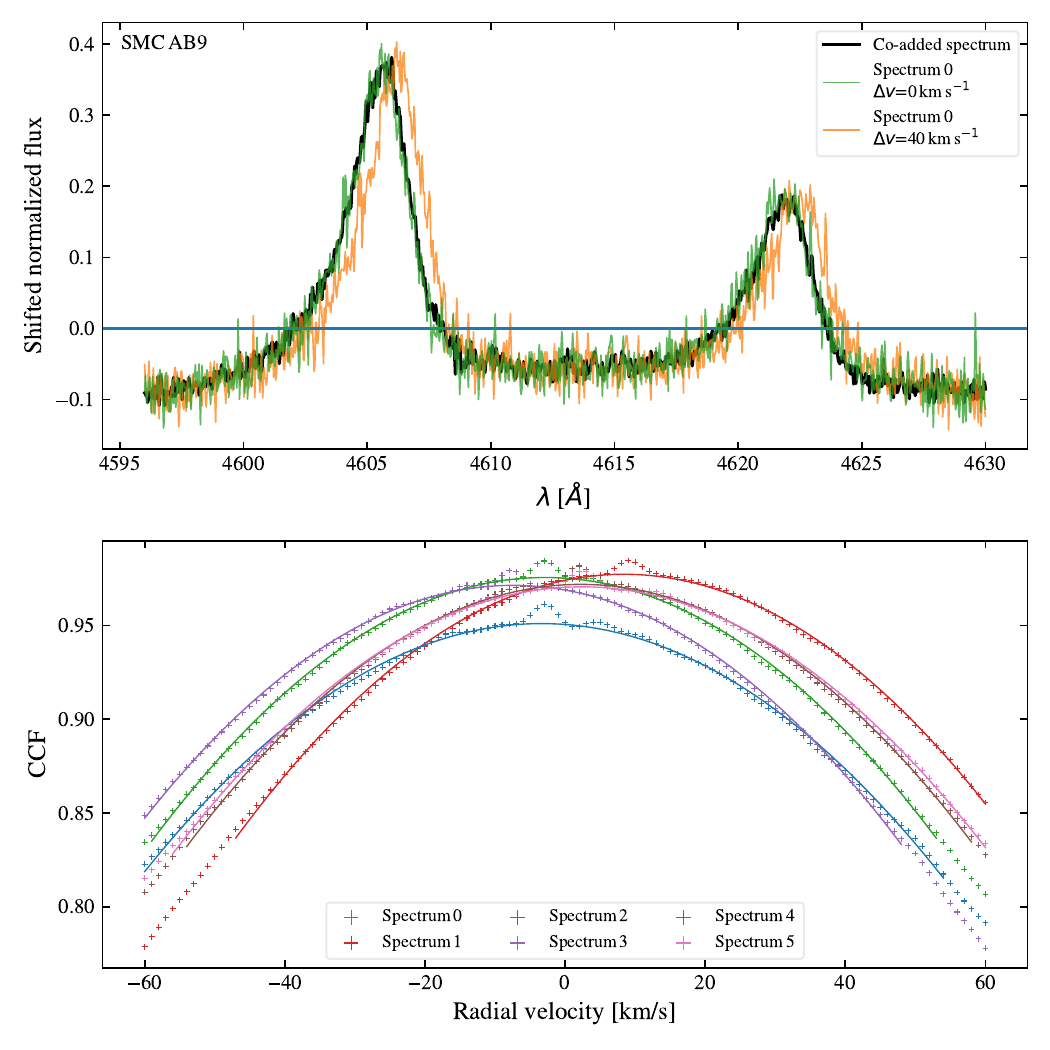} 
 \caption{\textit{Top:} The N\,{\sc v} 4604,\,4620\,\AA\ doublet that is used for cross-correlation, shown for SMC\,AB9. The co-added spectrum as well as the spectrum that was taken during observing night 0 are displayed. For reference, we also show spectrum\,0 shifted by 40\,km/s. The spectra are normalized and then vertically shifted such that the integrated flux in the shown wavelength range is zero.
 \textit{Bottom:} Cross-correlation functions (CCFs) of the N\,{\sc v} 4604,\,4620\,\AA\ doublet for the co-added template spectrum and the six individual spectra that were taken. 
 For this figure, the spectra were not corrected for barycentric motion.}
 \label{fig:ab9_xcorr}
\end{center}
\end{figure}

\begin{figure*}[t]
\begin{center}
\includegraphics[width=\linewidth]{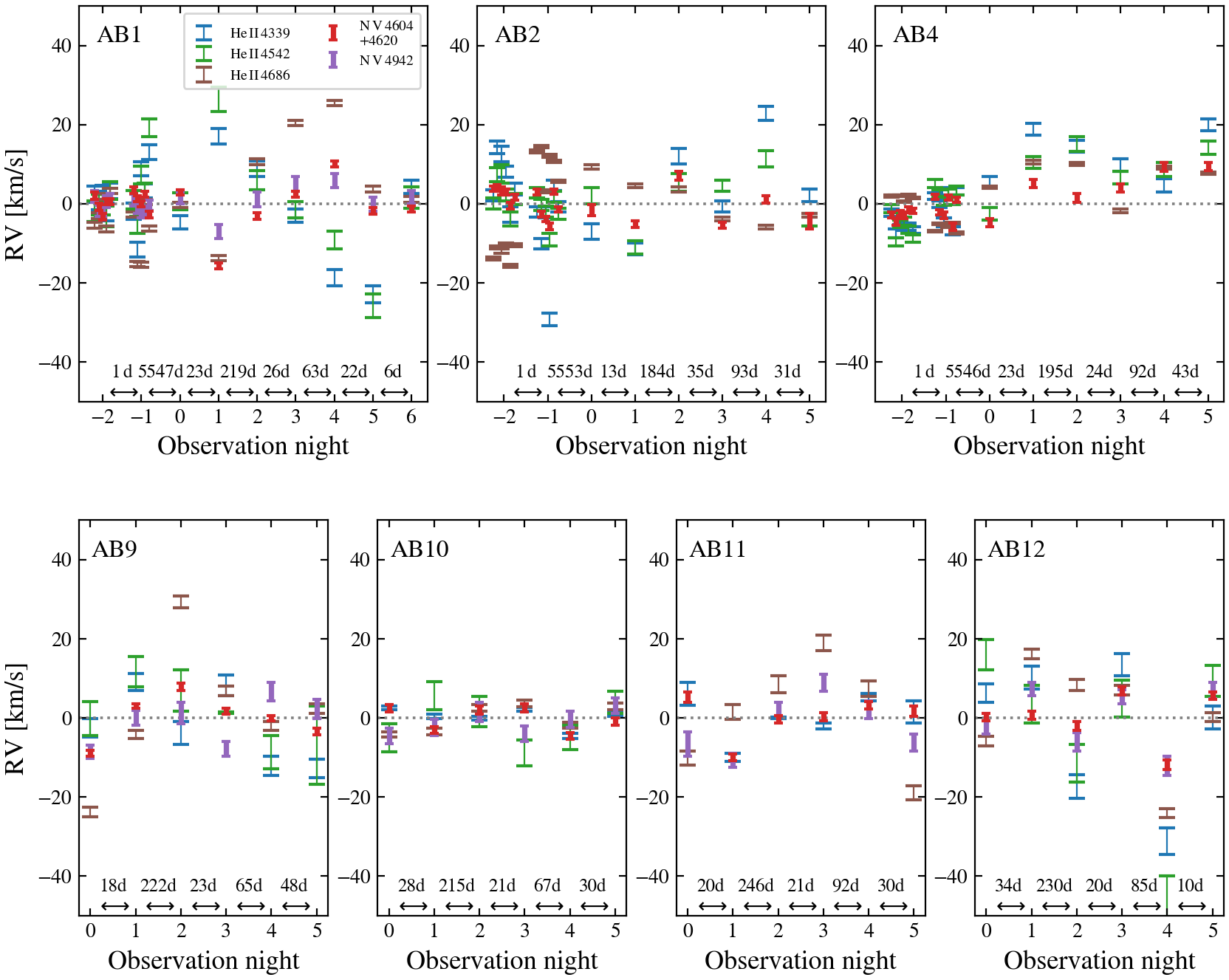} 
 \caption{RV measurements based on UVES-spectra taken during different nights. The first of the six spectra of our observing campaign is taken during observation night 0. The time interval in days between different observation nights is written at the bottom of the subplots. For SMC\,AB1, 2 and, 4, we also show measurements based on the pre-existing UVES-spectra of \cite{Marchenko07} (these are the data of observation nights -2 and -1). In this work we opt to use the N\,{\sc v} lines that are show with thicker lines.}
 \label{fig:t_vs_rv}
\end{center}
\end{figure*}

\subsection{Radial velocity measurements \label{sec:rv_measurements}}
We used a linear fit to the continuum to normalize our spectra (e.g., those shown in Fig.\,\ref{fig:ab9_lines}). 
First we selected a wavelength range around an absorption line of interest (e.g., 4596--4630\,\AA\ for the N$\,${\sc v} doublet shown in Fig.\,\ref{fig:ab9_xcorr}). We measured the average flux in the leftmost and rightmost 7.5\% of that wavelength range. Then, we divided the observed flux by a linear fit to the average flux in the central points of these two wavelength ranges.

Our method to measure the RV shifts of the normalized spectra is the cross-correlation method from \cite{Zucker03}. The same method has been successfully applied to precisely measure the RVs of Galactic WR stars \citep{Dsilva20, Dsilva22, Dsilva23}. We briefly describe the method below. First, the normalized spectra were shifted vertically such that the integrated flux is zero (Fig.\,\ref{fig:ab9_xcorr}, top). After that, for each of our WR stars we created a template spectrum by co-adding all available UVES spectra, starting with unshifted spectra. Then we iteratively shifted each individual spectrum by its measured RV (as described below) compared to the template spectrum, until the RVs no longer changed.
After that, we calculated the cross-correlation function (CCF) between the individual spectra and the template spectrum (see bottom panel of Fig.\,\ref{fig:ab9_xcorr}). We fitted a parabola to the CCF above 0.9 times its maximum value 
to obtain the best-fit RV and the statistical error $\sigma_\mathrm{CCF}$; the peak of the function fitted to the CCF is the best-fit RV value and the height and width of the fit to the CCF determine the value of the statistical error.

\section{Results \label{sec:results}}

Figure\,\ref{fig:t_vs_rv} shows our RV measurements for the apparently single SMC WR stars.
The time interval in days between two successive observations is written at the bottom of each panel. 

\subsection{Best lines for radial velocity measurements \label{sec:best_line}}
To decide on which line to use for our analysis, we show RVs from different lines in Fig.\,\ref{fig:t_vs_rv}. The N\,{\sc v} 4944\,\AA\ line is not observed in the relatively cold SMC WR stars AB\,2 and AB\,4, for which \cite{Hainich15} found $T_* \approx 45$\,kK, but it is detected in the spectra of the five other targeted SMC WR stars (where $T_*  \gtrsim 80$\,kK).

Observation night -2 and -1 in Fig.\,\ref{fig:t_vs_rv} refer to observations of SMC\,AB1, 2, and 4 of \cite{Marchenko07}, which are densely spaced during two successive nights. We notice that for these observations, the RVs measured with N\,{\sc v} lines (both the 4604\,+\,4620\,\AA\ doublet and the line around 4944\,\AA) remain relatively stable. The RVs measured with He\,{\sc ii} lines are not as constant. 
Also for the spectra taken in our recent campaign, the RVs measured from N\,{\sc v} lines are more stable than those from He\,{\sc ii} lines. 
For our seven target stars, we calculate $\Delta$RV (the difference between the highest and the lowest RV measurement) and $\sigma_\mathrm{RV}$ (the standard deviation of the RV measurements, where we subtract 1.5 degree of freedom in the divisor). We show the median values in Table\,\ref{tab:lines}. 
Compared to the measurements that use N\,{\sc v} lines, the median He\,{\sc ii} 4339\,\AA\ and He\,{\sc ii} 4542\,\AA\ $\Delta$RV measurements are $\sim$1.5 times higher; the measurements with the He\,{\sc ii} 4686\,\AA\ line yield a $\sim$2.5 times higher median $\Delta$RV.
Our interpretation for this is that the measurements based on He\,{\sc ii} are less accurate -- for example because these lines are more variable, weaker, or broader (Fig.\,\ref{fig:ab9_lines}). This is in agreement with findings of \cite{Dsilva22}. 
A possible explanation for lower variability is that N\,{\sc v} lines form in deeper layers than He\,{\sc ii} lines.

The cross-correlation method can also be applied to multiple line regions simultaneously. 
We did so for the N\,{\sc v} 4604\,+\,4620\,\AA\ doublet combined with N\,{\sc v} 4944\AA, as well as seven narrow He\,{\sc ii} and N features combined (see bottom two rows and caption of Table\,\ref{tab:lines}).
We find slightly more constant RVs for the combination of N\,{\sc v} lines than for the N\,{\sc v} 4604\,+\,4620\,\AA\ doublet individually. For the seven narrow He\,{\sc ii} and N\,{\sc v} features combined, we find the most constant RVs. 

For the RV measurements presented later on, we elect to use the RVs measured with the combination of the N\,{\sc v} 4604\,+\,4620\,\AA\ doublet and N\,{\sc v} 4944\AA\ line, because these are formed in the same relatively hot layers of the star that are close to the stellar surface. We consider this the conservative approach, since with the combined seven narrow He\,{\sc ii} and N\,{\sc v} lines we find the lowest RV variations.

\begin{table}[ht]
\caption{\label{tab:lines}
Median $\Delta$RV and $\sigma_\mathrm{RV}$ measured in our campaign for our seven target stars, using different lines.  
The narrow He\,{\sc ii} and N lines from the last row are He\,{\sc ii}\,4100, 4200, 4339, 4542\,\AA\ in combination with N\,{\sc iv}\,4058\,\AA\ and N\,{\sc v}\,4604+4620,  4944\,\AA.
}
\small
\centering
\begin{tabular}{lll}
\hline
\hline
 & $\Delta$RV$_\mathrm{median}$ & $\sigma_\mathrm{RV,\,median}$\\
Line (rest wavelength) & [km/s] & [km/s]\\
\hline
He\,{\sc ii} 4339\AA & 22 & 10.0 \\
He\,{\sc ii} 4542\AA & 22 & 8.3 \\
He\,{\sc ii} 4686\AA & 38 & 13.9 \\
N\,{\sc v} 4604+4620\AA & 15 & 5.6 \\
N\,{\sc v} 4944\AA (5/7 stars) & 15  & 6.3 \\
N\,{\sc v} 4604+4620\AA, 4944\AA & 14 & 5.4 \\
Narrow He\,{\sc ii} and N & 10 & 4.2 \\
\hline
\end{tabular}
\end{table}

\begin{table}[t]
\caption{\label{tab:summary_data}
Summary of the results of our RV monitoring campaign obtained using N\,{\sc v} 4604+4620, 4944 \AA\ lines. We describe in Sect.\,\ref{sec:abs_rv} how we obtain the absolute radial velocity $\mathrm{RV}_\mathrm{abs}.$}
\small
\centering
\begin{tabular}{llll}
\hline
\hline
Object & $\Delta$RV & $\sigma_\mathrm{RV}$ & RV$_\mathrm{abs}$  \\
 & [km/s] & [km/s] & [km/s]\\ 
\hline
SMC\,AB1 & 23.0 & 7.0 & $203 \pm 8$\\
SMC\,AB2 & 13.0 & 4.9 & $149 \pm 23$ \\
SMC\,AB4 & 14.3 & 5.2 & $144 \pm 23$ \\
SMC\,AB9 & 15.4 & 5.2 & $136 \pm 3$ \\
SMC\,AB10 & 5.5 & 2.4 & $152 \pm 25$ \\
SMC\,AB11 & 13.8 & 5.1 & $228 \pm 8$ \\
SMC\,AB12 & 18.4 & 6.9 & $156 \pm 22$\\
\hline
\end{tabular}
\end{table}

\subsection{Radial velocity variations \label{sec:rv_var}}

In search for binary reflex motion we consider the $\Delta$RV values that we measured for our individual WR stars.
We measure $\Delta$RV values between 6\,km/s and 23\,km/s (Table\,\ref{tab:summary_data}; individual measurements are tabulated in Appendix\,\ref{app:tabdata}).
These numbers are based on spectra taken during our observing campaign, but we note that RV measurements based on 15-year-old spectra of \cite{Marchenko07} fall within the RV range that we measure using the new spectra. 
The $\Delta$RV values that we measure are much lower than for the known SMC WR binaries, for which F03 measured values in the range $\Delta\mathrm{RV} = 400 - 600$\,km/s.

\begin{figure*}[t]
\begin{center}
\includegraphics[width=\linewidth]{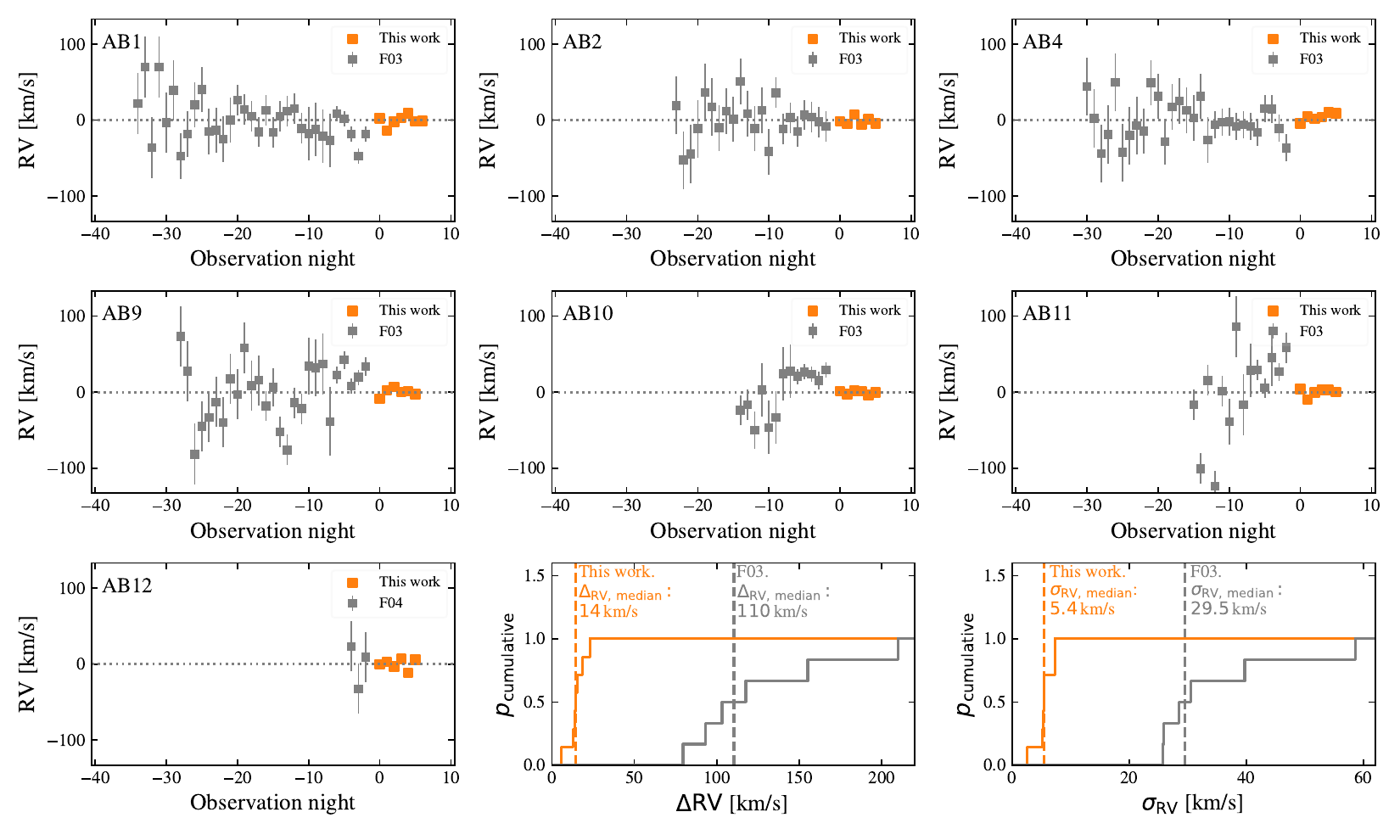} 
 \caption{ Similar to Fig.\,\ref{fig:t_vs_rv}, but showing RV measurements from a previous monitoring campaign (F03) and the three previous RV measurements of SMC\,AB12 from \cite{Foellmi04}. These previous measurements have been shifted by the previously measured average RV for each star. The center and right plots on the bottom row are cumulative probability distributions.
 }
 \label{fig:t_vs_rv_wf03}
\end{center}
\end{figure*}

We compare our $\Delta$RV values to earlier measurements of our targets. We use Engauge Digitizer\footnote{\url{http://markummitchell.github.io/engauge-digitizer/}} \citep{Mitchell20} to extract RV data from the plots of F03 and three measurements of SMC\,AB12 of \cite{Foellmi04}, as these were not tabulated. Figure\,\ref{fig:t_vs_rv_wf03} shows that we measure much more constant RVs for our target stars than F03 and \cite{Foellmi04} did. We further quantify the difference in the central lower panel of Fig.\,\ref{fig:t_vs_rv_wf03}. Our median $\Delta$RV value of 14\,km/s is a factor eight lower than in the earlier RV monitoring campaign, where the median $\Delta$RV value was 110\,km/s ($\Delta\mathrm{RV}$ range: $80 - 210$\,km/s).
This proves that at least the majority of the RV scatter seen by F03 is caused by measurement uncertainties rather than genuine RV motion.
Likely, multiple factors contribute to more precise measurements in our campaign. F03 measurements used the He\,{\sc ii} 4686\,\AA\ line, which we deemed to be less suitable than the N\,{\sc v} lines for RV measurements in Sect.\,\ref{sec:best_line}. Furthermore, our VLT-UVES spectra are of higher quality than those available for the previous campaign, in particular in terms of wavelength calibration (see sect.\,2.5 from F03) and resolving power ($R\sim40000$ vs. $R \sim 1000$). 

AB\,9 has been classified as a binary candidate by F03, whose measurements indicated a value of $\Delta \mathrm{RV} = 155$\,km/s for this source. These authors deemed it to be `marginally a binary from the RV point of view'. Their tentative orbital solution has $P_\mathrm{orb} = 37.6$\,d and radial velocity amplitude of $K = 43$\,km/s. Here we find that the RV of SMC\,AB9 is much more constant ($\Delta \mathrm{RV} = 15$km/s). This means that SMC\,AB9 was most likely a spurious detection in F03, as these authors already suspected. We note that when we use the He\,{\sc ii} 4686\,\AA\ line for our RV measurements of this source, we find $\Delta \mathrm{RV} = 53$km/s, making it likely that line profile variability \citep[LPV;][]{Fullerton96, Dsilva22} is at least partially responsible for the marginal detection in F03.

The median standard deviation of the RV measurements for individual WR stars is shown in the bottom right panel of Fig.\,\ref{fig:t_vs_rv_wf03}.
In the campaign of F03, the median $\sigma_\mathrm{RV} = 30$\,km/s lies somewhat above the 20\,km/s that they quote as typical error for their measurements.
Our campaign has a much lower median $\sigma_\mathrm{RV} \approx 5$\,km/s, in line with the $\Delta$RV result.
This $\sigma_\mathrm{RV} \approx 5$\,km/s is, however, larger than the statistical errors obtained from the fits to the CCFs, which are only of the order of 1\,km/s. The true error is most likely larger than that, for example because of LPV. 

In summary, we measure nearly constant RV values, which strongly reduces the possible binary parameter space for potential binary companions.
Still, we find some RV variability that could not be explained by the statistical errors alone. Below we explore two possibilities: that the small RV variability arises from LPV (Sect.\,\ref{sec:lpv}), and that our target stars are in binaries with low-mass and/or far-away companions (Sect.\,\ref{sec:possible_presence_of_companions}).

\section{Line profile variability \label{sec:lpv} }
\begin{figure}[t]
\begin{center}
\includegraphics[width=\linewidth]{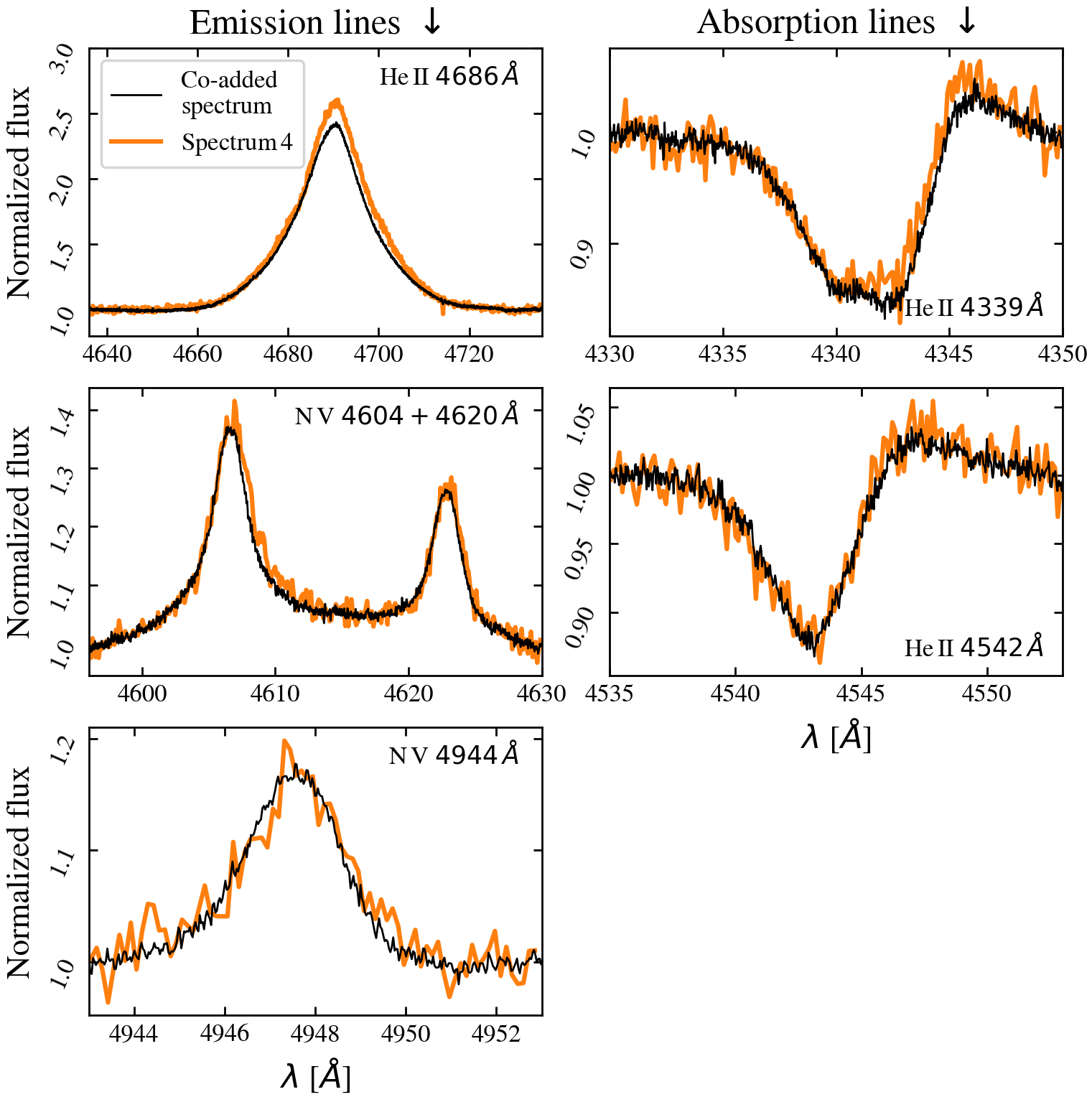} 
 \caption{Normalized cutouts of the co-added VLT-UVES spectrum of SMC\,AB1 (black lines).We also show the spectrum taken during observing night 4 in our observing campaign (orange lines), corrected for barycentric motion.}
 \label{fig:ab1_lpv}
\end{center}
\end{figure}

Here we take a more detailed look at spectrum\,4 of SMC\,AB1, taken during observation night 4, from which we infer an apparent antiphase RV motion between the N\,{\sc v} lines and the He\,{\sc ii} absorption lines (Fig.\,\ref{fig:t_vs_rv}). Figure\,\ref{fig:ab1_lpv} shows that LPV does take place in SMC\,AB1: compared to the co-added spectrum, spectrum\,4 exhibits a higher flux at the red side (i.e., at longer wavelengths) of the N\,{\sc v} and He\,{\sc ii} features. 
This excess of red flux causes an apparent shift to the red for the emission features (left panels of Fig.\,\ref{fig:ab1_lpv}), which lead us to measure positive values for their relative RVs. Conversely, the red excess causes an apparent shift to the blue for absorption features (right panels of Fig.\,\ref{fig:ab1_lpv}), and therefore we measured negative relative RVs for them. 

For the blue side of the N\,{\sc v} emission features and the He\,{\sc ii} absorption features in spectrum\,4, there seems to be no systematic shift compared to the co-added spectrum, which would be expected for RV offsets caused by binary motion. We conclude that LPV rather than binary motion is the most plausible explanation for the measured RV shifts in spectrum\,4 of SMC\,AB1, and for the apparent antiphase motion between absorption and emission lines.

This absorption/emission antiphase motion is not as clearly seen in all of the other spectra that we have taken. This could be explained by smaller variability, as we measured relatively large RV shifts in spectrum\,4 of SMC\,AB1. Furthermore, in SMC\,AB4 the He\,{\sc ii} 4339\,\AA\ and He\,{\sc ii} 4542\,\AA\ lines are in emission rather than in absorption, which may explain why the RV shifts measured with the He\,{\sc ii} and N\,{\sc v} lines tend to be in the same direction for individual spectra of this star (Fig.\,\ref{fig:t_vs_rv}).

A value of $\sigma_\mathrm{RV} \approx 5$\,km/s caused by LPV would be the same as what was found by \cite{Dsilva22}. These authors studied LPV from N\,{\sc v} lines by taking over 20 spectra in a narrow time interval of the wide binary star WR\,138 in the Milky Way, which is an early-type WN star, like our target stars.
We note that even larger RV variations (measured with a N\,{\sc v} line in spectra of WR\,6 in the Milky Way, a WN4 star) have been attributed to LPV caused by a co-rotating interaction region -- rather than binary motion -- by \cite{Barclay24}.

\section{Possible presence of undetected companions \label{sec:possible_presence_of_companions}}

\subsection{Sensitivity of our 
campaign
to binary motion
\label{sec:sensitivity}}

\begin{figure}[t]
\begin{center}
\includegraphics[width=\linewidth]{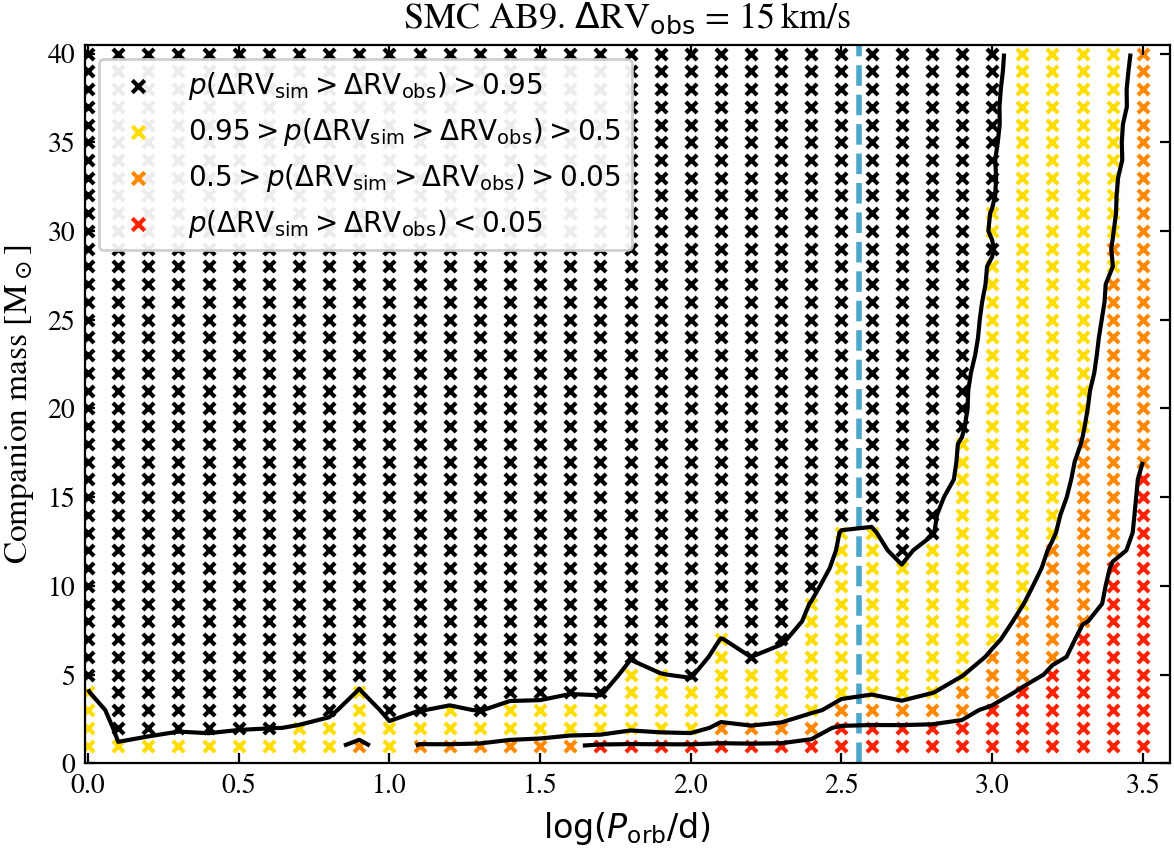} 
 \caption{
 Color-coded probability $p(\Delta \mathrm{RV}_\mathrm{sim}>\Delta \mathrm{RV}_\mathrm{obs})$ that binary-induced motion would lead to a RV variation exceeding the observed $\Delta \mathrm{RV}_\mathrm{obs}$ 
 in SMC\,AB9, for different combinations of orbital period $P_\mathrm{orb}$ and companion mass (see text for details).
 The higher this probability is, the less likely is the source an undetected binary. The dashed blue vertical line indicates $P_\mathrm{orb} = 1$\,yr. }
 \label{fig:pbinary_ab9}
\end{center}
\end{figure}

In order explore the allowed binary parameter space for companion stars, we follow a procedure similar to the one described in \cite{Dsilva20}, \cite{Dsilva22}, and \cite{Shenar23b}. We summarize the taken procedure below. 
We assume that our targeted WR stars have a mass of 20\,M$_\odot$ \citep{Hainich15, Schootemeijer18}.
We explore orbital periods in the range of $\log ( P_\mathrm{orb}/\mathrm{d} ) = [0, 0.1, \ldots, 3.5]$ and companion masses in the range of $ M_\mathrm{comp} = [1,2, \ldots, 40]$\,M$_\odot$ (see Fig.\,\ref{fig:pbinary_ab9}). For each of the 1440 combinations of orbital period and companion mass, we simulate our observing campaign 10000 times, by drawing a random orientation of the orbital plane (i.e., a random $\cos i$, where $i$ is the inclination angle to the line of sight) and a random orbital phase for the first observation. Then we measure the RV of the WR star in this mock binary at the same dates as in our observing campaign. 
We set the mock RV measurement error $\sigma_\mathrm{RV,\,mock}$ to zero. This is a conservative assumption, since we find that setting $\sigma_\mathrm{RV,\,mock} > 0$ increases, on average, the mock $\Delta$RV values. Then, for each companion mass and orbital period combination, we counted how often mock binary motion would result in a $\Delta$RV value larger than what is observed for the WR star.

\begin{figure}[t]
\begin{center}
\includegraphics[width=\linewidth]{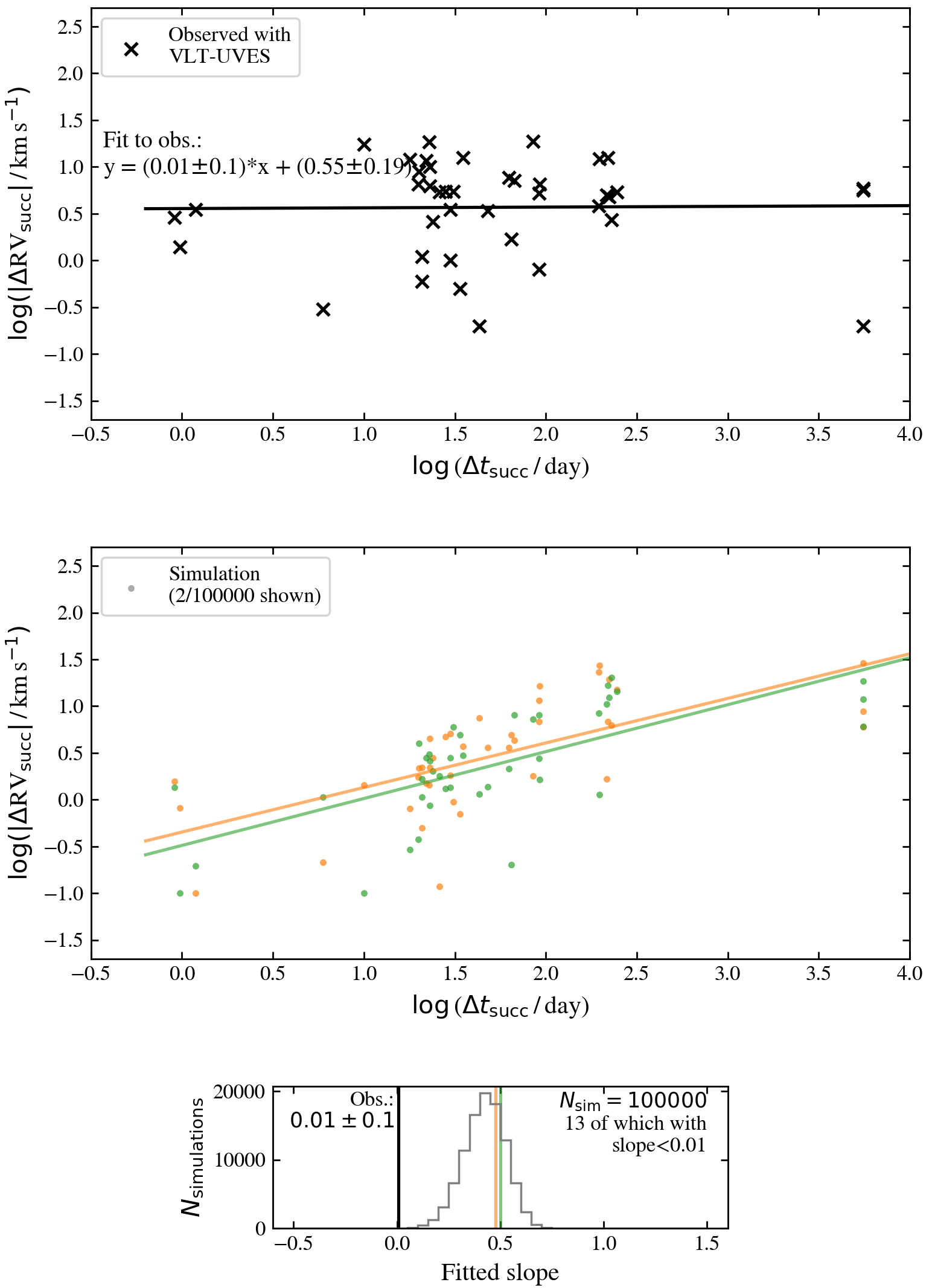} 
 \caption{\textit{Top:} 
 correlation of the time interval $\Delta t_\mathrm{succ}$ between two successive RV measurements and their observed $\Delta$RV$_\mathrm{succ}$.
 The black crosses represent measurements taken in this work and in \cite{Marchenko07}. \textit{Middle:} the colored dots represent measurements of a simulated population of long-period binaries with orbital periods between 1 and 10 years (see text for details).
 For clarity, we show only 2 out of 100000 simulated observing campaigns. \textit{Bottom:} histogram of the 100000 fitted slopes. The two vertical lines indicate the values of the two examples shown in the middle panel.
 }
 \label{fig:deltat_deltarv_planb}
\end{center}
\end{figure}

The results for SMC\,AB9 ($\Delta \mathrm{RV}_\mathrm{obs} = 15$\,km/s) are shown in Fig.\,\ref{fig:pbinary_ab9}. 
This figure shows that we would have most likely measured higher $\Delta$RV values than observed if SMC\,AB9 had a binary companion. For companion masses above 5\,M$_\odot$ and orbital periods below one year, this probability exceeds 95\% in 98,6\% of the parameter space.

We can obtain an overall probability $\hat{p}$ that binarity would have been detected by our measurements when we specify underlying orbital period and mass ratio probability distributions. For flat distributions of $\log (P_\mathrm{orb} / d)$ and mass ratio, this overall probability is just the average of the probability values displayed in Fig.\,\ref{fig:pbinary_ab9}, which becomes $\hat{p} = 0.90$. However, binary evolution models predict that in a scenario where the WR star formed through binary stripping, configurations with very low-mass companions, as well as very short or very long orbital periods, are avoided. When using underlying orbital period and mass ratio distributions motivated by binary population synthesis models \citep[][see Appendix\,\ref{sec:pint} for details]{Renzo19,Langer20} we find even higher values of $ \hat{p}> 0.97$, while when we truncate these distributions for $P_\mathrm{orb} > 30$\,d  -- since those systems have already been detected -- we still find $\hat{p}>0.94$ for SMC\,AB9.

For our other target stars we obtain similar results (Figs.\,\ref{fig:pdetect_ab1}-\ref{fig:pdetect_ab12}; Table\,\ref{tab:integrated_pdetect}). 
Hence, for each individual WR star we can rule out most of the realistic parameter space for binaries that may have previously interacted, in particular where $P_\mathrm{orb} < 1$\,yr. We discuss binaries with $1\,\mathrm{yr} < P_\mathrm{orb}< 10\,\mathrm{yr}$ in Sect.\,\ref{sec:long_p}.
Performing the same experiment with the F03 data reveals the possibility that companions of some tens of Solar masses with $P_\mathrm{orb} \approx 1$\,yr could have been missed (Figs.\,\ref{fig:pdetect_ab1_f03}-\ref{fig:pdetect_ab12_f03}; Table\,\ref{tab:integrated_pdetect_f03}).

We also compute the most likely companion mass of our WR stars under the assumption that, apart from the small $\sigma_\mathrm{CCF} \approx 1$\,km/s, the measured RV variability is caused by binary motion.  Figure\,\ref{fig:rvmatch_ab9} shows that for SMC\,AB9, this is about $1-3$\,M$_\odot$ for periods of up to one year, increasing to beyond 5\,M$_\odot$ for periods longer than 3\,yr. The most likely companion masses reside very close to the 0.5 probability contour in Fig\,\ref{fig:pbinary_ab9}.

To construct Fig.\,\ref{fig:pbinary_ab9} we assumed circular orbits, since we expect that if binary interaction has taken place this would have circularized the orbit \citep[][eq.\,20]{Iorio23}.
As a test, we have repeated the simulations described in this section with a flat eccentricity distribution in the range $0 \leq e \leq 0.9$. This lead to detection probabilities that were smaller but only slightly so (Table\,\ref{tab:integrated_pdetect_ecc} and Fig.\,\ref{fig:pbinary_ab9_flatecc}), and do therefore not change the interpretation of our results.

For all the considered ranges of the possible binary parameter space described in Appendix\,\ref{sec:appa}, we find the lowest values for $\hat{p}$ when excluding orbital periods below 30\,d and including eccentric orbits (Table\,\ref{tab:integrated_pdetect_ecc}). 
For the seven target WR stars, we then find an average of $\hat{p}_\mathrm{P30+}$ = 0.801. This implies a chance that we missed a binary companion for \textit{all} of our target WR stars of the order of $(1-0.801)^7 \approx 10^{-5}$ even for this set of assumptions.

\subsection{Long-period binaries \label{sec:long_p}}

In the last section we have seen that our small observed $\Delta$RV values make it unlikely that our target stars have binary companions, especially with orbital periods of a year or less. Because we observed during three successive semesters, it could be more likely that we missed binary companions with an orbital period of the order of a few years in our campaign. 

Here we perform an additional experiment where we explore the likelihood that \textit{all} our target WR stars are in 1-10\,yr binaries. 
If so, we would expect to measure small RV differences for observations that are separated days or weeks apart, and larger RV differences for observations taken months or years apart.
In the top panel of Fig.\,\ref{fig:deltat_deltarv_planb} we plot with black crosses the time between successive observations ($\Delta t_\mathrm{succ}$) against the RV difference between them ($\Delta \mathrm{RV}_\mathrm{succ}$). For our seven target stars, Fig.\,\ref{fig:deltat_deltarv_planb} shows a total of 42 data points for successive VLT-UVES observations, including both our observations and those from \cite{Marchenko07}.
There is no significant correlation in the observational data, as we obtain a value of $0.01 \pm 0.10$ for the slope of the linear fit ( also shown in the top panel of Fig.\,\ref{fig:deltat_deltarv_planb}). 
This in line with our measured $\sigma_\mathrm{RV} \approx 5$\,km/s being caused by a combination of measurement errors and LPV on timescales around one day or shorter.

To quantify how likely it is that $\Delta t_\mathrm{succ}$ and $\Delta \mathrm{RV}_\mathrm{succ}$ do not show a significant positive correlation if all our target stars are in $1-10$\,yr binaries, we again simulate our observing campaign as in Sect.\,\ref{sec:sensitivity}, but now 100000 times, for all seven target stars. We draw a random logarithm of the orbital period in the range $0 \leq \log ( P_\mathrm{orb} / \mathrm{yr} ) \leq 1 $ and a random mass ratio in the range $ 0.05 \leq M_\mathrm{comp} / M_\mathrm{WR} \leq 2$ (i.e., a companion mass in the range $1 - 40$\,M$_\odot$), both from flat distributions \citep{Opik24, Kouwenhoven07}. We add a Gaussian error to each mock binary measurement, based on the observed $\sigma_\mathrm{CCF}$.
The green and orange dots in the middle panel of Fig.\,\ref{fig:deltat_deltarv_planb} show the results for two individual simulated observing campaigns; the other 99998 are omitted for visibility purposes. We find that in all except 13 simulations (99.987\%), the linear fit of $\Delta t_\mathrm{succ}$ and $\Delta \mathrm{RV}_\mathrm{succ}$ is larger than the observed value of 0.01 (bottom panel of Fig.\,\ref{fig:deltat_deltarv_planb}). Therefore, it is highly unlikely that all our target stars are long-period binaries with of $P_\mathrm{orb} = 1-10$\,yr. 

\subsection{Neutron star companions \label{sec:nss}}
Neutron stars (NSs) have masses in the range 1-2\,M$_\odot$, and hence we could only detect them as companions to our target WR stars if they had orbital periods of $\sim$30 days or less (Fig.\ref{fig:pbinary_ab9}). However, NSs are expected to emit X-rays arising from the accretion of material from the wind of the WR stars. F03 found upper limits of $ L_\mathrm{x} < 1-2\cdot 10^{33}$\,erg/s based on ROSAT data. Using eq.\,1 from \cite{Shapiro76} for the NS accretion radii and the stellar parameters from \cite{Hainich15}, and adopting Bondi-Hoyle accretion, we find that these X-ray non-detections exclude the presence of NSs with orbital periods of $P_\mathrm{orb} \lesssim 100$\,d. However,
the accretion rates, and thus the X-ray flux,  could be significantly smaller if the magnetosphere of the rotating NS star would expel a fraction of the inflowing wind material.

Independent of this, it appears unlikely that all the apparently single WR stars have NS companions, for three reasons. First, the observed WR stars in the SMC are thought to originate from main-sequence stars with $M \gtrsim 40$\,M$_\odot$ \citep{Schootemeijer18, Shenar20}. Any more evolved companion, e.g., a compact star, would have an initial mass similar to that or higher, and therefore would most likely be a BH and not an NS \citep{Schneider23}. Second, the majority of NS producing binaries are expected to break up when the corresponding SN occurs \citep{Eldridge11, Renzo19}, which makes it unlikely that the number of WR+NS binaries is comparable to the number of WR+O star progenitors \citep[of which there are five in the SMC;][]{Shenar16}. Third, binary interaction in massive non-disrupted systems with NS companions has been claimed to always lead to a merger, explaining the absence of detected WR$+$NS binaries \citep{vandenHeuvel17, Toala18}.
Consequently, it appears unlikely for each of our WR stars to have an NS companion, and we dismiss the possibility that \textit{all} our target WR stars have NS companions.

\section{Other signatures of binarity \label{sec:other_signatures}}

\subsection{Photometric variability}
We analyze photometric data that has become available in the last 20 years. F03 have investigated time-series photometry data of all of the SMC WR stars but SMC\,AB12 and the eclipsing binary SMC\,AB5 \citep{Moffat98}. They found no periodic signals except for the apparently single star SMC\,AB4, for which they reported a 3.2$\sigma$ detection of a 6.55\,d period variability. This detection was for MACHO-B data \citep{Alcock99} but there were non-detections for MACHO-R and OGLE I data \citep{Udalski98}.

Using the {\sc lightkurve} python package \citep{Lightkurve18}, we analyze OGLE III \citep{Udalski08} and OGLE IV \citep{Udalski15} time-series photometry data of all seven of our targets stars (A. Udalski, private communication). These data cover a timespan of about twenty years. With the `Lomb-Scargle' method we obtain the power spectrum for each target WR star, exploring $10^6$ equidistant frequencies in the range $0.001\, \mathrm{d}^{-1} < f < 0.67\,\mathrm{d}^{-1}$. Then we used the {\tt flatten} function to calculate the SNR spectrum. 
We provide an overview of the OGLE data and the highest SNR values calculated from them in Table\,\ref{tab:ogle}.
Nowhere in the considered frequency range did any of our target stars approach a threshold S/N of 4.6 \citep{Bowman21} for periodic variability. 
We also provide the power spectra and phase-folded light curves in Fig.\,\ref{fig:periodograms}, which further support the absence of convincing binary-induced signals.
We are thus not able to find a significant periodic signal for any of our target stars, including SMC\,AB4.

\subsection{Imprints of companions in the spectra}
SMC\,AB1, 9, 10, 11, and 12 are hot objects \citep[$T_* \geq 80$\,kK;][]{Hainich15} and therefore potential cooler companions could be found by their imprints in the spectra.
For example, we look for the He\,{\sc i} 4471\,\AA\ line, which appears in stars later than O5 \citep{Walborn00}. None of the hot WR stars in our sample show any sign of this He\,{\sc i} line (also not after rebinning). This might be unsurprising since even in the composite spectra of the five binary SMC WR stars \citep[which have O-star companions of tens of Solar masses; see][]{Shenar16} the He\,{\sc i} 4471\,\AA\ absorption line does not go deeper than $0.02 - 0.1$ times the continuum. For the presence of lower-mass (and thus dimmer) stars, our RV measurements can be expected to place much stronger constraints than line strength measurements.

Other lines that we attempt to find in search for companions of hot WR stars are He\,{\sc i} 4387\,\AA, He\,{\sc i} 4713\,\AA, He\,{\sc i} 4922\,\AA, Si\,{\sc iv} 4089\,\AA, Si\,{\sc iv} 4116\,\AA, and O\,{\sc ii} 4350\,\AA. We note that these tend to be weaker than the He\,{\sc i} 4471\,\AA\ line in the spectra of WR binaries in the Magellanic Clouds \citep{Shenar16, Shenar19}. We do not see a signature of any of these lines in the spectra of the hot WR stars that we targeted.
We conclude that we are unable to find imprints of companion stars in the WR spectra, in line with the results from our RV measurements.

Finally, we inspect spectra of SMC\,AB2 and SMC\,AB4 from the Ultraviolet Legacy Library of Young Stars as Essential Standards \citep[ULLYSES;][]{Roman-Duval20}. We notice that for both stars, the C\,{\sc iv} feature around 1550\,\AA\ is saturated, which is indicative of an absence of another hot star (because another hot star with a weaker wind than the WR star could contribute to the flux in this line region). The only other SMC WR star in the ULYSSES sample is SMC\,AB9, which is much hotter than SMC\,AB2 and SMC\,AB4 and lacks a strong C\,{\sc iv} feature.

\subsection{Absolute radial velocities and proper motions \label{sec:abs_rv}}

So far we have found little evidence for our target WR stars currently being in binary systems. In principle, they may have been in binary systems in the past, that have broken up or merged (see also Sect.\,\ref{sec:formation}). If they were in binary systems that have been disrupted by supernova explosions, this will result in higher space velocities. To test this, we investigate their absolute RVs and proper motions.

\begin{figure}[t]
\begin{center}
\includegraphics[width=\linewidth]{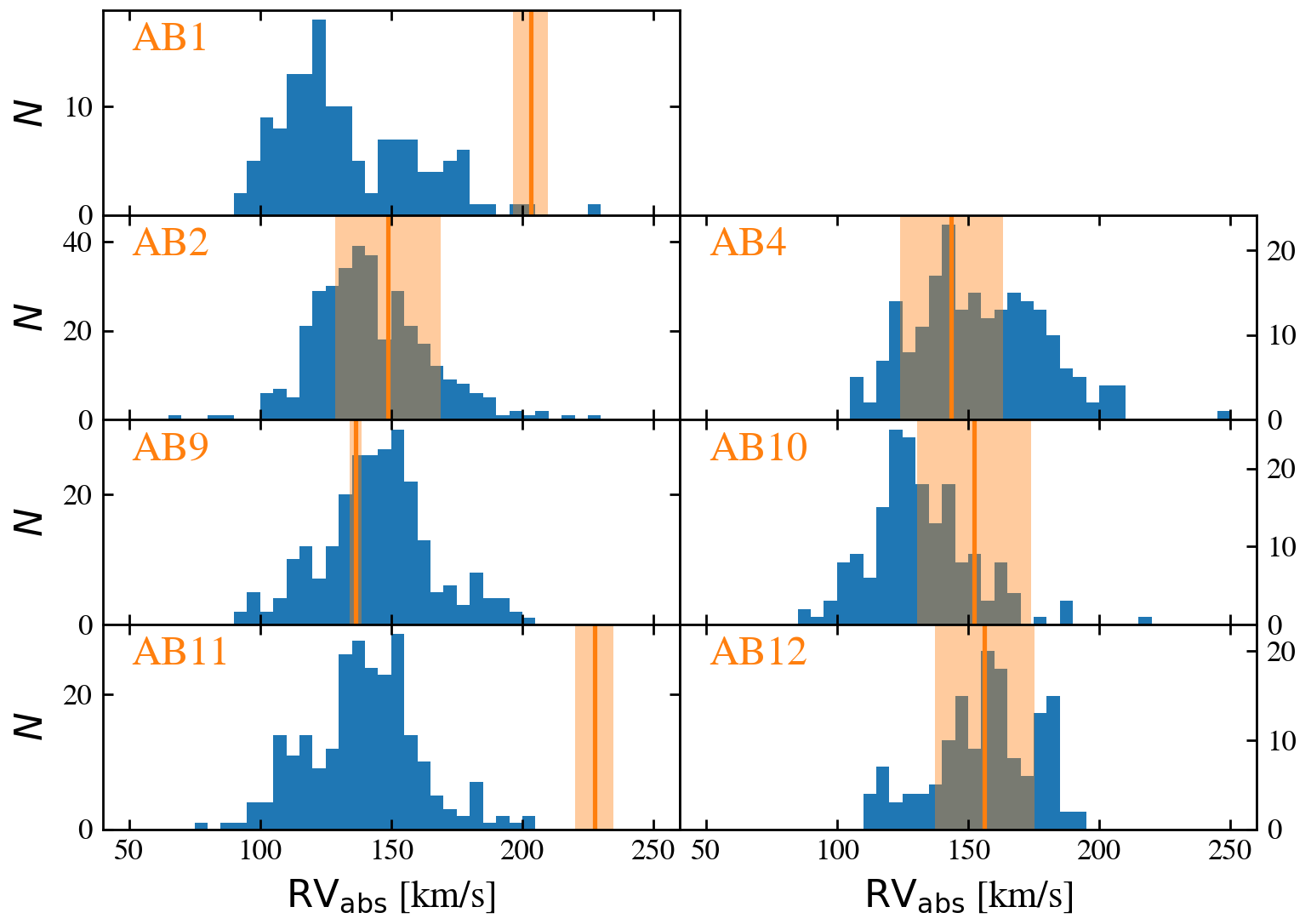} 
 \caption{ RVs of our targeted WR stars (vertical orange lines; the shaded region indicates the error margin). The blue histograms show RVs of GAIA DR3 sources in a 15\arcmin radius.
 }
 \label{fig:abs_rvs}
\end{center}
\end{figure}

The absolute RVs are obtained from cross-correlating our co-added spectra (Sect.\,\ref{sec:rv_measurements}) with the best-fit model atmosphere spectra from \cite{Hainich15}. 
Based on where the best fit of this model to the observations is achieved, we select three line regions for the cross-correlation. Two of these are the He\,{\sc ii} 4339\,\AA\ and He\,{\sc ii} 4686\,\AA\ lines. As the third line region we select N{\,\sc iv} 4058\AA\ for the two coldest objects (SMC AB\,2 and SMC\,AB4), and N{\,\sc v} 4944\,\AA\ for the other five WR stars.
We define our measured absolute RV as the average measured RV and the error as the standard deviation of the three measurements (again with 1.5 degree of freedom subtracted in the divisor). 
We correct the co-added spectra for barycentric motion by subtracting the average barycentric velocity, which we obtain with the {\tt baryCorr} function of the python package \textsc{pyasl}. \citep{Piskunov02}
The results are shown in Table\,\ref{tab:summary_data}. We then compare the absolute RVs of our WR stars to those from stars in their vicinity. For this comparison, from the GAIA archive\footnote{\url{https://gea.esac.esa.int/archive/}} we download all DR3 sources \citep{GAIA23} that have measured RVs and reside within 15\arcmin\ of our target WR stars. We exclude sources with a parallax larger than five times the parallax error to remove foreground sources \citep{Aadland18, Schootemeijer21}. Fig.\,\ref{fig:abs_rvs} shows the absolute RVs of our targeted WR stars and the nearby sources within 15\arcmin. 

Five out of seven single WR stars have absolute RVs that are consistent with those of sources in their surroundings. For SMC\,AB11, its $\mathrm{RV_\mathrm{abs}}$ of $228 \pm 8$\,km/s is significantly different to sources in its surroundings, where we find $\mathrm{RV_\mathrm{abs} = 139 \pm 21}$\,km/s (the error being the standard deviation of the nearby sources' RV, and 139\,km/s being their average RV value). This makes SMC\,AB11 a runaway star, moving with a relative velocity of $91\pm23$\,km/s in the line-of-sight direction. Similarly, for SMC\,AB1 we derive a relative velocity of $68 \pm 35$\,km/s based on its $\mathrm{RV_\mathrm{abs} = 203 \pm 8}$\,km/s, and an $\mathrm{RV_\mathrm{abs} = 135 \pm 35}$\,km/s for sources in its vicinity. This qualifies SMC\,AB1 also as a runaway star. 

\begin{figure}[t]
\begin{center}
\includegraphics[width=\linewidth]{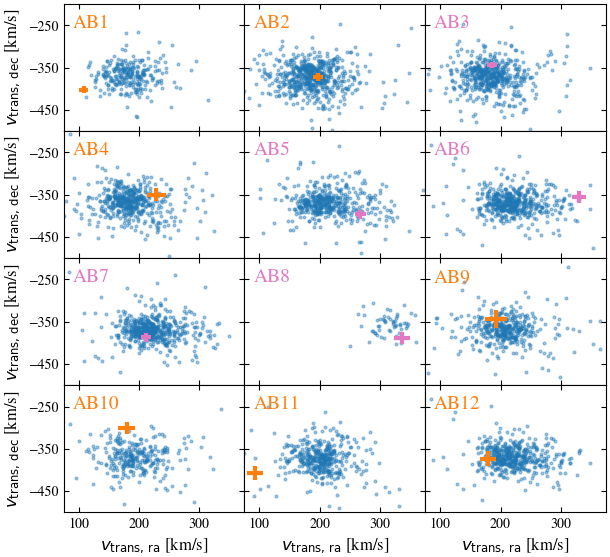} 
 \caption{Velocities transverse to the line of sight. Our targeted WR stars are shown in orange, and binary SMC WR stars are shown in magenta. The shown errors are obtained from proper motion errors. GAIA DR3 sources in a 5\arcmin\ radius are shown in blue.
 }
 \label{fig:vtrans}
\end{center}
\end{figure}

As a check, we also look up the proper motions of SMC WR stars in GAIA DR3 and compare them to surrounding sources within 5\arcmin. We remove foreground sources as described above, and we discard sources dimmer than $G = 17$ or with $\mathrm{RUWE} > 1.4$. We translate the proper motions intro transverse velocities $v_\mathrm{trans}$ adopting an SMC distance of 62.44\,kpc \citep{Graczyk20}. The results are shown in Fig.\,\ref{fig:vtrans}. 
The transverse  velocities of most of our WR stars are similar to those of their surrounding sources, except, again, for SMC\,AB1 and SMC\,AB11. 
We find a relative two-dimensional $v_\mathrm{trans} \approx 107\pm33$\,km/s for SMC\,AB11 compared to the average of the surrounding stars (the error follows from the GAIA proper motion uncertainty of SMC\,AB11). This confirms its nature as a runaway star. 
Similarly, for SMC\,AB1 we measure a relative $v_\mathrm{trans} = 78$\,km/s, in line with the result for the absolute RV.

Puzzlingly, we also find a high relative transverse velocity for SMC\,AB6, which is unexpected given that this is part of a multiple system \citep{Shenar18}. This might indicate that the SMC WR proper motions are more uncertain than the errors indicate, perhaps because these very massive stars that tend to be in crowded regions. Additionally, we note that, given the errors, our method is not sensitive to intermediate-velocity runaway stars.

We take a closer look at the runaway star SMC\,AB11,
which has a bright red source only 1\arcsec\ away that we also put in the slit (Sect.\,\ref{sec:obs}).
From the composite spectrum, we measure the absolute RV of the red source to see if it moves in the opposite direction of SMC \,AB11, which would be expected if they were in a multiple system that has been broken up. However, using the narrow H\,$\alpha$ line we find 
$\mathrm{RV}_\mathrm{abs} \approx 150$\,km/s, which is a very typical value in its environment (Fig.\,\ref{fig:abs_rvs}). This implies that the red source is a chance alignment rather than that it originates from a common multiple system with SMC\,AB11.

Finally, we investigate the supernova remnant (SNR) population in the SMC \citep{Maggi19, Matsuura22}. SMC \,AB11 is closest to SNR J0052-7236, which resides at a projected distance of 50\,pc (3\arcmin). It has an age of about 35\,kyr \citep{Leahy22}. SMC\,AB11 would need much higher velocity of $v_\mathrm{trans} \approx 1500$\,km/s to reach a 3\arcmin\ separation within the lifetime of the SNR. In addition, the proper motion vector does not point back to SNR J0052-7236. Therefore, we cannot associate SMC\,AB11 to an SNR. We conclude the same for SMC\,AB1, which is not close to any known SNR in the SMC.
Our inability to link SMC\,AB1 and SMC\,AB11 to a `smoking gun' SNR does not rule out that they are post-supernova runaways, since the helium-burning WR lifetime is about ten times longer than the oldest SNR in the SMC \citep[which is 35\,kr;][]{Leahy22}. 
We will further discuss the implications of the runaway nature of SMC\,AB1 and SMC\,AB11 in Sect.\,\ref{sec:formation}.

\section{Discussion \label{sec:discussion} }

\cite{Schootemeijer18} found a fundamental difference in the internal constitution between binary and apparently single WR stars in the SMC. The binary WR stars showed a shallow internal radial H/He-gradient as expected from a retreating convective core during central hydrogen burning, whereas the apparently single WR stars were shown to have an up to 10-times steeper H/He gradient, and a correspondingly less massive H-rich envelope. The shallow gradient of the WR stars in binaries is well reproduced by the mass transferring short period binary evolution models of \cite{Pauli22}. However, the steep H/He-gradients require another WR star formation history.

To form a steep H/He-gradient requires that the region with the naturally shallow gradient is mixed. As discussed by \cite{Schootemeijer18}, this
may occur either by allowing the WR progenitor to expand during its post main-sequence evolution, such that convection induced by either hydrogen shell burning or by a convective envelope could mix the corresponding layers, or by increasing the convective core in a hydrogen-burning mass gainer. In both cases, the WR star might have lost its envelope by binary mass transfer or, as a single star, by winds or eruptions. Below, we discuss the scenarios proposed in the literature that allow for the formation of single WR stars.

\subsection{Formation channel of single WR stars in the SMC \label{sec:formation}}

\paragraph{Chemically homogeneous evolution.}
In CHE models, rotationally-induced mixing brings hydrogen-depleted material to the surface of exceptionally rapidly spinning stars \citep[e.g.,][]{Maeder87, Brott11}. This scenario works most efficiently at low metallicity \citep{Yoon06}. As such, CHE could in principle explain the presence of hydrogen-poor WR stars in the SMC \citep{Martins09, Hainich15, Ramachandran19}; and see \cite{Koenigsberger14} for SMC\,AB5. However, five out of seven of the single WR stars (SMC\,AB1, 9, 10, 11, and 12) are too hot to be explained by core hydrogen burning models \citep[][their fig.\,2]{Schootemeijer18}, but have, at the same time, a high surface hydrogen mass fraction \citep[0.36 on average;][]{Hainich15}. Their chemical structure is thus very inhomogeneous. In addition, the apparently single SMC WR stars show low projected rotational velocities \citep{Martins09, Hainich15, Vink17}. 

\paragraph{Stellar mergers.}
We expect that mergers involving H-rich stars do not produce merger products resembling the observed single SMC WR stars, as these typically contain less than a Solar mass of hydrogen \citep{Schootemeijer18}. In principle it is possible to produce a H-poor merger product when both stars develop He cores and then interact. Such an evolutionary path has been produced for the 2\,M$_\odot$ quasi-WR star HD\,45166 \citep{Shenar23}. However, at high stellar masses (where stars do not develop degenerate helium cores, meaning that companions have little time to catch up in their evolution)
this might require fine-tuning of the initial mass ratio and physics assumptions \citep{Pols94, Wellstein99}.
Also, if mergers of two H-poor stars produced all single SMC WR stars, we would expect that in the WR+MS systems, which are their progenitors, the MS stars are close to the end of their H-burning lifetime or close to filling their Roche lobes. However, typically the MS companions are early O stars that are far from filling their Roche lobes \citep{Shenar16}, which are not expected to exhaust H in their cores or otherwise instigate interaction before the end of the life of their companion. We are therefore unaware of a merger channel that is likely to produce seven out of twelve SMC WR stars.

\paragraph{Binary stripping.}
At first glace it seems that binary companions have not stripped our seven targeted SMC WR stars, because it is unlikely that we missed them (Sect.\,\ref{sec:possible_presence_of_companions}). While it is possible that we missed some rather low mass companion in a wide orbit, it is implausible that such a star could have removed the hydrogen envelope of a WR star in the past. As the progenitors of our WR stars were more massive than $\sim$40\,M$_\odot$ (Sect.\,\ref{sec:nss}), their binaries with companion masses below $\sim$5\,M$_\odot$ would have such an extreme initial mass ratio that stable mass transfer is not expected \citep{Henneco23}. Unstable mass transfer in a system with such a mass ratio is predicted to result in a merger \citep{Kruckow16}. 

It is, in principle, possible that a WR star is stripped by a binary companion that has since exploded as a supernova, which disrupted the binary system. However, as discussed for the merger scenario above, this requires one binary component to end hydrogen burning before the other reaches core collapse. This is unlikely to happen in most of the initial binary parameter space. 
Still, binary stripping might be a minority channel -- see also the cases of SMC\,AB1 and SMC\,AB11 (Sect.\,\ref{sec:abs_rv}).

Recently, \cite{Pauli23} explained an apparent bimodality in the temperatures of SMC WR stars by proposing that the colder WR stars SMC\,AB 2 and 4 are accretors in binary systems. 
This is in contrast to the lack of detected RV variability in our study.

\paragraph{Single star mass loss.}

Evolutionary models with SMC or near-SMC metallicity tend to use wind mass loss recipes in which winds are too weak to strip stars with initial masses around 40\,M$_\odot$ \citep[e.g.,][]{Brott11, Georgy13, Choi16, Limongi18, Schootemeijer19}. Could the mass loss rates have been underestimated in these models? For MS stars, the models typically use the recipe of \cite{Vink01} for which recently evidence has built up that it overestimates rather than underestimates MS mass loss \citep[by a factor $\sim$3;][]{Bjorklund21, Brands22}.
Perhaps more likely is that stars lose a significant amount of mass 
as cool stars after their core-hydrogen burning phase. 

One possibility for post-main-sequence self-stripping is eruptive mass loss in luminous blue variable (LBV) stars. 
The LBV phenomenon might be related to stars reaching the Eddington limit \citep{Grafener12, Sanyal17}. 
According to recent predictions, this can happen in evolved SMC stars with initial masses as low as 30\,M$_\odot$ and lead to eruptive mass loss that can strip their H-rich envelope \citep{Cheng24}. This seems to be consistent with the initial masses of $M\gtrsim30-40$\,M$_\odot$ for SMC WR stars (Table\,\ref{tab:mini_mfin}).
The SMC harbors two stars that show or have shown LBV behavior \citep[see][also for two more candidates]{Richardson18}. One of these (HD\,5980) underwent an LBV outburst in the 1990s and at present manifests itself as a WR star in the only WR+WR binary in the SMC \citep{Koenigsberger10}. The other is R\,40 \citep[][]{Szeifert93, Agliozzo21}, which is currently in an LBV eruption episode, and it has an uncertain but seemingly high mass loss rate of $10^{-4} - 10^{-3}$\,M$_\odot$/yr \citep{Campagnolo18}. Both sources imply that LBV mass loss could play a role in the production of WR stars at low metallicity.
An SMC population of two LBVs and of the order of ten wind-stripped WR stars would be in line with the LMC and Milky Way, where LBVs are also more rare than WR stars by a factor of at least a few \citep{Richardson18, vanderHucht01, Shenar19}. 
A caveat might be that LBVs in the LMC seem to be much more isolated than O and WN (although not WC) stars, which could indicate that (at least in the LMC) LBVs could be accretors in binary systems \citep{Smith15}.

Wind mass loss during a red supergiant (RSG) phase could also be responsible for stripping the outer layers of our WR progenitors. Mass loss rates of RSGs are notoriously uncertain and also their metallicity dependence is not established. Therefore, RSG mass loss rates in the SMC could occupy a wide range of values 
 -- see e.g. fig.\,15 of \cite{Yang23}. 
Some mass loss rate predictions shown there increase steeply with luminosity and reach values at $\log( L / \mathrm{L}_\odot ) \approx 5.5$ that are high enough to strip H-rich envelopes of massive stars on timescales of $\sim$10\,kyr \citep{Beasor23, Yang23}. However, these mass-loss rates at the bright end are the most uncertain, where RSGs are scarce.
Related to this, we note that \cite{Kee21} found that high and metallicity-independent mass loss rates can be achieved if turbulent velocities in RSGs are high.

\paragraph{Resum\'e.}
From the discussion above, we conclude that  chemically homogeneous evolution, stellar mergers, and binary stripping are unlikely to constitute the dominant channel for producing single WR stars in the SMC. It follows that single star mass loss -- most probably in the LBV or RSG phase -- is likely responsible for the loss of the H-rich envelope in the majority of the investigated SMC WR stars.

\begin{figure*}[t]
\begin{center}
\includegraphics[width=0.80\linewidth]{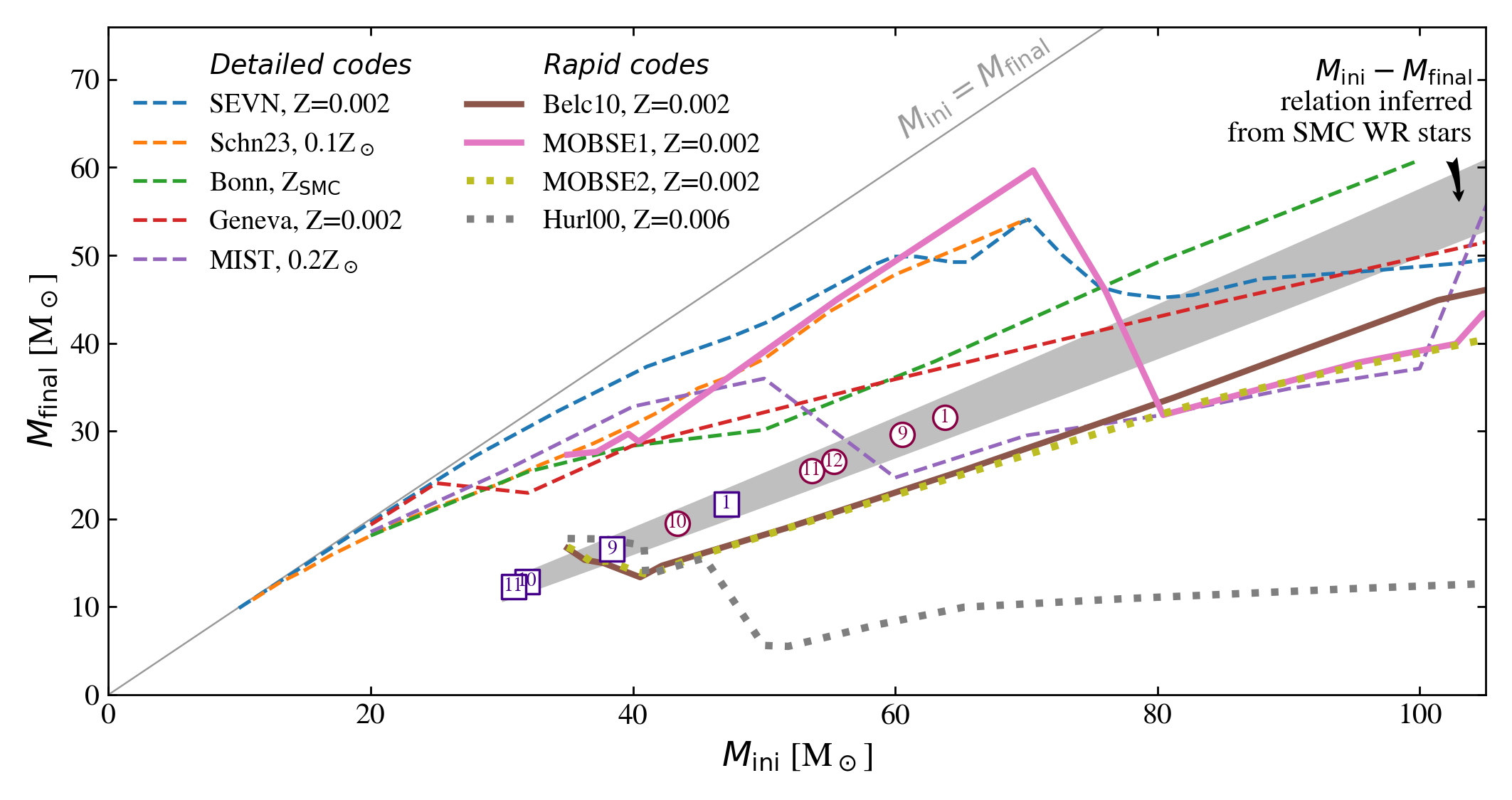} 
 \caption{Initial mass -- final mass relations for SMC stars. Data points of SMC WR stars are shown with circles \citep[based on][]{Hainich15} and squares \citep[based on][]{Martins09, Martins23}, the numbers indicating their identifier (e.g., 1 for SMC\,AB1). 
 The grey shaded region indicates He core masses at the terminal-age main sequence ($M_\mathrm{He\,core,\,TAMS}$)  from evolutionary models of \cite{Schootemeijer19}, for overshooting values between 0.22 and 0.44. The colored lines show the following relations from literature: 
 SEVN \citep{Spera15}, Schn23 \citep{Schneider23}, Bonn \citep[][]{Schootemeijer19}, Geneva \citep{Georgy13}, MIST \citep{Dotter16, Choi16}, Belc10 \citep[][data taken from fig.\,2 of \cite{Banerjee20}]{Belczynski10}, MOBSE \citep{Giacobbo18}, and Hurl00 \citep[][data taken from fig.\,1 of \cite{Belczynski10}]{Hurley00}.
 }
 \label{fig:mini_mfinal}
\end{center}
\end{figure*}

\subsection{Implications for the outcome of massive binary interaction at low metallicity.}

After showing that binary evolution was likely not important  for the majority of the seven targeted SMC WR stars, let us assume that the detected binaries constitute the whole SMC WR binary population. 
All five binaries have short orbital periods in the range $6\,\mathrm{d} < P_\mathrm{orb} < 20\,\mathrm{d}$. 
We use initial-final orbital separation relations \citep{Soberman97, Tauris06} to find possible solutions for the initial orbital periods for SMC\,AB6, 7, and 8 \citep[SMC\,AB3 has a poorly defined mass ratio, and the initial-final separation formula likely does not apply to the evolutionary history of SMC\,AB5; see][]{Koenigsberger14, Schootemeijer18}. 
Under the assumption that mass stripped away from the donor either ends up on the accretor or is lost from the accretor, there are no solutions for $P_\mathrm{orb, \, ini} > 25$\,d, regardless of the adopted mass transfer efficiency.
Even though this is low-number statistics, the apparent absence of medium- to long-period binaries in the SMC WR population appears striking. It could imply that massive longer-period binaries at low metallicity merge upon binary interaction.
This might impede the channel for producing binary BH mergers from isolated binary evolution, which often involves stars with large initial periods \citep[][]{Belczynski16, Kruckow18}. Therefore, this needs to be carefully checked in future work.

Interestingly, there are also no known binaries with $P_\mathrm{orb} \geq 40$\,d among the `classical' WR stars on the nitrogen sequence (subtype earlier than WN5) in the Large Magellanic Cloud \citep[LMC --][]{Foellmi03b, Shenar19}.
Since there are over ten times more WN stars in the LMC than in the SMC, this seems to exacerbate the lack of long- and intermediate period WR binaries at low metallicity.  However, the LMC WR stars have not been monitored with high-quality observations such as those presented in this study. The LMC would thus be a good place to further investigate this issue.

Additionally, all SMC WR stars in binaries have companions that are of similar or higher mass than the WR star \citep[with mass ratios between 1 and 3; F03,][]{Shenar16}. This may require a non-negligible mass transfer efficiency in the systems that survived binary interaction.

\subsection{Implications for BH masses at low metallicity \label{sec:BH_masses}}

Our finding that the majority of the apparently single SMC WR stars lost their envelope without the help of a binary companion has direct consequences for the BH population produced by the massive stars in the SMC. It makes these stars uniquely suitable to investigate the initial mass -- final mass ($M_\mathrm{ini} -M_\mathrm{final}$) relation because of two properties: i) they have H-rich layers near their surfaces, but these contain only $\sim$1\,M$_\odot$ of material \citep{Schootemeijer18}, and, ii), they have low mass loss rates \citep[at which they lose $M \lesssim 1\,\mathrm{M}_\odot$ during their WR lifetime;][]{Hainich15}. This implies that their current mass, their He core mass, and their final mass $M_\mathrm{final}$ are practically the same\footnote{SMC\,AB2 and SMC\,AB4 are relatively cold objects that could also burn H in their cores \citep{Schootemeijer18}. We therefore exclude them from this analysis.}.
We will use this property to obtain the $M_\mathrm{ini} - M_\mathrm{final}$ relation of the single SMC WR stars.

First, we obtain the He core masses of the SMC WR stars from the best-fit models  of \cite{Schootemeijer18}, from their large grid of synthetic massive star models with pre-defined helium core masses, surface hydrogen abundances, and H/He gradients. 
We then use the initial mass -- He core mass relations from stellar evolution models \citep[][with an overshooting value of 0.33]{Schootemeijer19} to obtain $M_\mathrm{ini} -M_\mathrm{final}$ relations. 
To estimate the errors, we assume that the uncertainty in luminosity \citep[from][]{Hainich15, Shenar16, Shenar18} propagates into the uncertainty in mass. The error on the mass then depends on the slope of the mass-luminosity relation, which we obtain from \cite{Grafener11}. We repeat the analysis based on the observational WR star properties obtained by \cite{Martins09} and \cite{Martins23}, who found somewhat lower luminosities.

The results define the empirical initial-final mass relation shown as the thick grey line in Fig.\,\ref{fig:mini_mfinal}, whose width reflects the error in this relation (see also Tab.\,\ref{tab:mini_mfin}). It is well defined in the initial mass range $40 \dots 65$\,M$_\odot$ (or $30 \dots 50$\,M$_\odot$ based on the luminosities of Martins et al.), and extrapolated to $\sim 100$\,M$_\odot$.

In Fig.\,\ref{fig:mini_mfinal}, we compare our result with initial-final mass relation or initial mass-BH mass relations obtained from detailed stellar evolution calculations, or used in so-called rapid binary evolution models. To obtain data from works from the literature that provide no online data apart from their plots, we used Engauge Digitizer again. \cite{Belczynski10}, \cite{Banerjee20}, and \cite{Giacobbo18} provide remnant masses rather than final stellar masses --
for these, to obtain the final masses we divide the remnant masses by 0.90, as these studies assumed that the gravitational BH mass is 10\% smaller than the baryonic mass. 
We only show the mass range $M_\mathrm{ini} > 32$\,M$_\odot$, since for lower masses, where the CO core mass is lower than 11\,M$_\odot$, these codes assume BH fallback fractions lower than unity \citep{Fryer12}.
The $M_\mathrm{ini} - M_\mathrm{final}$ relations from the literature roughly come in two flavors: those that experience strong LBV mass loss in the mass range $40 < M_\mathrm{ini} / \mathrm{M}_\odot < 80$, and those that do not.

\begin{table}[t]
\caption{\label{tab:mini_mfin}
Initial masses and He core masses found for the SMC WR stars. The errors on $M_\mathrm{core}$ and $M_\mathrm{ini}$ are correlated. \textit{Potsdam} refers to \cite{Hainich15} for the single stars and \cite{Shenar16} for the binaries. \textit{Martins} refers to \cite{Martins09} and \cite{Martins23}. Notes have the following meaning. $^\mathrm{a)}$: object has a surface temperature that matches both H-core and He-core burning models; $^\mathrm{b)}$: values from \cite{Wang19b}.}
\small
\centering
\begin{tabular}{lllll} 
\hline 
 \hline
 & $M_\mathrm{core} / \mathrm{M}_\odot$ & $M_\mathrm{ini} / \mathrm{M}_\odot$ & $M_\mathrm{core} / \mathrm{M}_\odot$ & $M_\mathrm{ini} / \mathrm{M}_\odot$ \\

\textit{Based on:} & \textit{Potsdam} & \textit{Potsdam} & \textit{Martins} & \textit{Martins} \\
\hline
\textit{Single:} & & & & \\
SMC\,AB1 & 31.6$^{+11.4}_{-8.3}$ & 63.8$^{+18.7}_{-14.1}$ & 21.7$^{+5.1}_{-4.2}$ & 47.1$^{+8.7}_{-7.1}$ \\
SMC\,AB2$^{a)}$ & 16.3$^{+2.4}_{-2.1}$ & 37.9$^{+4.1}_{-3.6}$ & 13.3$^{+1.9}_{-1.7}$ & 32.6$^{+3.3}_{-2.9}$ \\
SMC\,AB4$^{a)}$ & 18.4$^{+2.8}_{-2.4}$ & 41.4$^{+4.8}_{-4.2}$ & 21.3$^{+3.4}_{-2.9}$ & 46.5$^{+5.8}_{-5.0}$ \\
SMC\,AB9 & 29.6$^{+10.6}_{-7.8}$ & 60.5$^{+17.5}_{-13.4}$ & 16.6$^{+3.9}_{-3.1}$ & 38.4$^{+6.7}_{-5.5}$ \\
SMC\,AB10 & 19.5$^{+6.2}_{-4.7}$ & 43.4$^{+10.6}_{-8.2}$ & 12.9$^{+2.9}_{-2.4}$ & 31.9$^{+5.0}_{-4.1}$ \\
SMC\,AB11 & 25.5$^{+8.6}_{-6.4}$ & 53.6$^{+14.4}_{-11.0}$ & 12.3$^{+2.7}_{-2.2}$ & 30.9$^{+4.8}_{-3.9}$\\
SMC\,AB12 & 26.5$^{+9.1}_{-6.8}$ & 55.3$^{+15.2}_{-11.5}$ & & \\
\textit{Binary:} & & & & \\
SMC\,AB3 & 24.7$^{+1.9}_{-1.8}$ & 52.2$^{+3.2}_{-3.0}$ & & \\
SMC\,AB5$^{a)}$ & 43.1$^{+7.8}_{-6.8}$ & 82.8$^{+12.6}_{-10.8}$ & & \\
SMC\,AB6 & 22.9$^{+3.6}_{-3.1}$ & 49.2$^{+6.1}_{-5.3}$ & & \\
SMC\,AB7 & 30.4$^{+5.1}_{-4.4}$ & 61.8$^{+8.5}_{-7.3}$ & & \\
SMC\,AB8$^{b)}$ & 35 & 80 & & \\
\hline
\end{tabular}
\end{table}

\subsubsection{Initial -- final mass relations with strong LBV mass loss}

The mass-loss prescription from \cite{Belczynski10}
includes a strong, metallicity-independent LBV mass loss rate of $1.5 \cdot 10^{-4}$\,M$_\odot$/yr, which in practice removes all the hydrogen-rich layers of stars born more massive than 40\,M$_\odot$. It is widely used to predict BH-BH merger rates in rapid binary evolution codes \citep[e.g., COMPAS;][]{Riley22}. It is also used in a non-default simulation of MOBSE \citep[MOBSE2;][]{Giacobbo18} and the triple evolution code TRES \citep{Toonen16, Kummer23}. Also many N-body simulations of dynamical BH-BH binary formation include the prescription of \cite{Belczynski10}, for example simulations of young star clusters \citep{Chattopadhyay22, Fragione22} which include NBODY7 \citep{Banerjee20, Banerjee21} and PeTar \citep{WangIwasawa20, Barber23}, as well as simulations around galactic nuclei \citep{AntoniniRasio16, Tagawa20, Fragione22b}, and simulations of globular clusters \citep{Rodriguez16, AntoniniGieles20}. 
The high amount of late-phase mass loss in these simulations seems to be in agreement with our findings at SMC metallicity. However, we find (while assuming negligible WR mass loss) somewhat higher final masses than what is predicted by \cite{Belczynski10}, most likely because of their high WR mass loss rates, which are about five times larger than those that are observed for single SMC WR stars \citep{Hainich15}.

The mass loss prescription of \cite{Hurley00} includes, apart from strong LBV mass loss, metallicity-independent WR winds. These overestimate the mass loss rates of SMC WR stars by about a factor twenty compared to what is observed by \cite{Hainich15}, and lead to very low final masses. The \cite{Hurley00} prescription is used in BH-BH merger simulations of globular clusters \citep{Askar17, Hong18, Samsing18} and star clusters near galactic nuclei \citep{Petrovich17}. 

\subsubsection{Initial -- final mass relations with weak LBV mass loss}

Many codes use or obtain $M_\mathrm{ini} -M_\mathrm{final}$ relations under the assumption of less drastic late-phase mass loss near SMC metallicity, which leads to higher final masses than we infer for SMC WR stars.
Examples are SEVN \citep{Spera15}, which uses PARSEC models \citep{Tang14, Chen15}, the single-star models of \cite{Schneider23}, MIST \citep{Dotter16, Choi16}, Geneva models \citep{Georgy13}, and Bonn models \citep[here showing][which uses the same mass-loss recipe as \cite{Brott11} and ComBinE \citep{Kruckow18}]{Schootemeijer19}. 
Moreover, in the latest prescription of MOBSE \citep{Giacobbo18} called MOBSE1, the LBV mass loss prescription is $Z$-dependent for stars with initial masses up to $\sim$80\,M$_\odot$ \citep[i.e., up to the point where stars approach the Eddington limit, following][]{Tang14, Chen15}. This allows 40-80\,M$_\odot$ models near SMC metallicity to retain more H-rich layers and produce much more massive BHs than with the prescription of \cite{Belczynski10}. The MOBSE1 prescription is used in other rapid binary evolution codes \citep[e.g., COSMIC;][]{Breivik20}
and N-body simulations of young star clusters \citep{DiCarlo19, Santoliquido20, Rastello21, Torniamenti22}, nuclear star clusters \citep{Stephan19, Wang21}, and globular clusters \citep{Mapelli21}.
Our results imply that the mass loss rates used in these calculations are 
mostly too low. 

\subsubsection{Implications for population synthesis of double BH merger progenitors}
Our results indicate that $M_\mathrm{ini} -M_\mathrm{final}$ relations that include strong late-phase mass loss \citep[e.g.,][]{Belczynski10} are preferable at $Z_\mathrm{SMC}$. Relations calculated without strong late-phase mass loss around SMC metallicity in the initial mass range of 40\,M$_\odot$ to 80\,M$_\odot$ (e.g., MOBSE1, SEVN) seem to overestimate final masses of single stars. 
This result might be of particular importance for simulations of double BH mergers produced via the dynamical channel, since there stars that lived as single stars can be involved in double BH mergers.

For double BH merger progenitors produced by isolated binary evolution, the inferred strong late-phase mass loss might only be of importance if it is LBV rather than RSG mass loss that sheds the envelope. Unlike strong RSG mass loss, strong LBV mass loss might prevent very massive stars from expanding to large radii, and thereby prevent a common-envelope phase that tightens the BH-BH progenitor system. Our study cannot identify which of the two channels (LBV or RSG mass loss) is favorable. 

\section{Conclusions \label{sec:conclusions}}
We have used high-quality spectra taken with VLT-UVES during three consecutive semesters to monitor the RVs of all seven apparently single SMC WR stars, in search of binary reflex motion. To measure the RVs we used narrow N\,{\sc v} lines, which turned out to be remarkably stable. We found $\sigma_\mathrm{RV}$ values of $\sim$5\,km/s and all seven WR stars had $\Delta$RV values in the range of 6\,km/s to 23 km/s,
including the former binary candidate WR star SMC\,AB9. 
We argue that LPV from atmospheric or wind variability can account for these small RV variations.

For individual stars, our Monte Carlo simulations exclude the presence of companions more massive than 5\,M$_\odot$ and orbital periods shorter than one year with $\sim$95\% confidence or more. Statistically, this makes it practically impossible that we missed such companions for all of them. 
Companions with smaller masses are thought to rather merge with their primary star than to strip its envelope.
When considering underlying orbital period distributions up to ten years, we still estimate a probability below $\sim$10$^{-5}$ that all our seven target SMC WR stars are binaries.
Based on a lack of correlation of $\Delta$RV measurements between successive exposures and the time elapsed between them,
we can further rule out that all of our target stars are in long-period ($1-10$\,yr) binaries. Moreover, our simulations show that previous studies were not sensitive to a large part of the binary parameter space. The fraction of SMC WR stars with detected binary companions remains at 0.42.

SMC\,AB1 and SMC\,AB11 appear to be runaway stars with relative line-of-sight velocities of $\sim$80\,km/s, compared to their surroundings. The proper motions of these sources strengthen this finding. While such high velocities are not preferred in population synthesis models, \cite{Eldridge11} predict runaway WR stars with space velocities up to 120\,km/s. Furthermore, \cite{Evans11} and \cite{Sana22} find runaway O\,stars with up to  $\sim$90\,M$_\odot$in the LMC with similarly high runaway velocities.

Our results imply that for SMC stars above $\sim$40\,M$_\odot$, single stars lose the major part of their hydrogen-rich envelope, and that a binary companion is not required for this. We argue that late-phase mass loss in the LBV or RSG phase is the most likely mechanism to strip these stars so they become WR stars. Our results put strong constraints on the maximum black hole mass which can emerge from stars above $\sim$40\,M$_\odot$.
 
All five binary WR stars in the SMC show orbital periods below 20\,d. Although their number is small, this could imply that massive interacting long-period binaries at low metallicity merge. While more work is needed to substantiate this hypothesis, it would suggest that the predictions for the double black hole merger rate based on isolated evolution of long period binaries are largely overestimated.

\acknowledgements{
We thank the anonymous referee for a constructive report and useful suggestions.
We warmly thank Andrzej Udalski for sending us the OGLE-III and OGLE-IV data of our target stars. Equally warmly, we thank Rainer Hainich for providing the best-fit model atmosphere spectra. AS thanks Frank Backs, Sarah Brands, Michal Pawlak, Philipp Podsiadlowski, and Fabian Schneider for useful discussions. We thank ESO for delivering the excellent spectra that made this study possible.
The research leading to these results has received funding from the European Research Council (ERC) under the European Union's Horizon 2020 research and innovation programme (grant agreement numbers 772225: MULTIPLES).
}

\bibliography{bib}{}
\bibliographystyle{aa}

\clearpage

\appendix

\section{Observational data \label{app:tabdata}}

\begin{table}[ht]
\caption{\label{tab:ab1_data}
RV measurements of SMC\,AB\,1. We provide the RVs relative to the co-added spectrum obtained using the N\,{\sc v} 4604+4620\,\AA\ and N\,{\sc v} 4944\,\AA\ lines. The absolute RVs can be found in Table\,\ref{tab:summary_data}. Measurements before MJD 54000 are based on spectra from \cite{Marchenko07}. The MJD 
is measured at the start of the observations. }
\small
\centering
\begin{tabular}{lll}
\hline \hline
MJD & RV$_\mathrm{N\,V}$ & $\sigma_\mathrm{CCF, \, NV}$\\
\hline
53974.08143819 & -2.8 & 0.8 \\
53974.15281245 & 0.6 & 0.6 \\
53974.22478018 & 0.6 & 0.6 \\
53974.29469711 & 2.4 & 0.8 \\
53974.3652008 & -1.3 & 0.8 \\
53975.06148374 & 3.1 & 0.8 \\
53975.13130698 & 1.4 & 0.7 \\
53975.20126692 & -2.2 & 0.7 \\
53975.27429798 & -0.5 & 0.7 \\
53975.34441279 & 0.8 & 0.7 \\
59522.241237232 & 2.7 & 0.6 \\
59545.140053771 & -13.8 & 0.7 \\
59764.337670124 & -2.1 & 0.8 \\
59790.286486529 & 2.9 & 0.7 \\
59853.155227237 & 9.1 & 0.7 \\
59875.139356245 & -1.4 & 0.7 \\
59881.104598668 & -0.9 & 0.7 \\
\hline
\end{tabular}
\end{table}

\begin{table}[ht]
\caption{\label{tab:ab2_data}
Same as Table\,\ref{tab:ab1_data}, but for SMC\,AB2.}
\small
\centering
\begin{tabular}{lll}
\hline \hline
MJD & RV$_\mathrm{N\,V}$ & $\sigma_\mathrm{CCF, \, NV}$\\
\hline
53974.05813941 & -0.6 & 0.8 \\
53974.12974506 & 3.9 & 0.6 \\
53974.20138833 & 1.7 & 0.8 \\
53974.27175485 & 3.0 & 0.7 \\
53974.341774 & 4.0 & 0.8 \\
53974.41218787 & 2.9 & 1.0 \\
53975.03795807 & -5.6 & 0.8 \\
53975.1080113 & -2.4 & 0.7 \\
53975.17794324 & 2.9 & 0.7 \\
53975.24800434 & 3.1 & 0.7 \\
53975.32088677 & -1.2 & 0.7 \\
53975.390976 & -3.8 & 0.7 \\
59528.098763616 & -4.7 & 0.9 \\
59541.075943645 & -1.5 & 1.4 \\
59725.389460631 & 7.5 & 1.2 \\
59760.400719872 & -5.5 & 0.8 \\
59853.198151712 & 1.0 & 1.0 \\
59884.112963287 & -4.4 & 1.9 \\
\hline
\end{tabular}
\end{table}

\begin{table}[ht]
\caption{\label{tab:ab4_data}
Same as Table\,\ref{tab:ab1_data}, but for SMC\,AB4.}
\small
\centering
\begin{tabular}{lll}
\hline \hline
MJD & RV$_\mathrm{N\,V}$ & $\sigma_\mathrm{CCF, \, NV}$\\
\hline
53974.04129595 & -4.7 & 0.7 \\
53974.11267138 & -1.7 & 0.7 \\
53974.18478773 & -1.2 & 0.6 \\
53974.25548055 & -2.4 & 0.7 \\
53974.32562773 & -2.9 & 0.7 \\
53974.39570418 & -2.7 & 0.8 \\
53975.02142447 & -6.2 & 0.9 \\
53975.09172646 & -3.2 & 0.8 \\
53975.16159462 & -2.1 & 0.6 \\
53975.23155571 & 1.3 & 0.7 \\
53975.30485427 & 1.2 & 0.7 \\
53975.3747245 & 0.8 & 0.6 \\
59521.267478467 & -4.2 & 1.0 \\
59544.223161156 & 4.6 & 1.1 \\
59739.40864374 & 1.7 & 1.1 \\
59763.390003412 & 3.8 & 1.0 \\
59855.126683198 & 10.1 & 1.2 \\
59898.151501907 & 8.8 & 1.0 \\
\hline
\end{tabular}
\end{table}

\begin{table}[ht]
\caption{\label{tab:ab9_data}
Same as Table\,\ref{tab:ab1_data}, but for SMC\,AB9. }
\small
\centering
\begin{tabular}{lll}
\hline \hline
MJD & RV$_\mathrm{N\,V}$ & $\sigma_\mathrm{CCF, \, NV}$\\
\hline
59527.235134485 & -8.6 & 0.6 \\
59545.182260154 & 2.7 & 0.6 \\
59767.263809198 & 6.8 & 1.0 \\
59790.329875426 & 0.1 & 0.7 \\
59855.137487622 & 1.1 & 0.8 \\
59903.169638222 & -2.5 & 0.7 \\
\hline
\end{tabular}
\end{table}

\begin{table}[ht]
\caption{\label{tab:ab10_data}
Same as Table\,\ref{tab:ab1_data}, but for SMC\,AB10. }
\small
\centering
\begin{tabular}{lll}
\hline \hline
MJD & RV$_\mathrm{N\,V}$ & $\sigma_\mathrm{CCF, \, NV}$\\
\hline
59524.15053773 & 1.3 & 0.8 \\
59552.13830278 & -2.9 & 0.8 \\
59767.310841186 & 1.9 & 0.9 \\
59788.20105391 & 1.4 & 1.0 \\
59855.18150306 & -3.7 & 0.8 \\
59885.044768909 & -0.4 & 0.9 \\
\hline
\end{tabular}
\end{table}

\begin{table}[ht]
\caption{\label{tab:ab11_data}
Same as Table\,\ref{tab:ab1_data}, but for SMC\,AB11.}
\small
\centering
\begin{tabular}{lll}
\hline \hline
MJD & RV$_\mathrm{N\,V}$ & $\sigma_\mathrm{CCF, \, NV}$\\
\hline
59524.210034441 & 4.0 & 1.2 \\
59544.171211109 & -9.8 & 0.8 \\
59790.372793215 & -0.6 & 0.7 \\
59811.253413358 & 2.9 & 0.9 \\
59903.113894512 & 2.8 & 0.9 \\
59933.070686762 & 0.1 & 1.1 \\
\hline
\end{tabular}
\end{table}

\begin{table}[ht]
\caption{\label{tab:ab12_data}
Same as Table\,\ref{tab:ab1_data}, but for SMC\,AB12.}
\small
\centering
\begin{tabular}{lll}
\hline \hline
MJD & RV$_\mathrm{N\,V}$ & $\sigma_\mathrm{CCF, \, NV}$\\
\hline
59527.176912331 & -0.6 & 0.8 \\
59561.038740174 & 3.1 & 0.9 \\
59791.223275872 & -3.1 & 1.0 \\
59811.306473575 & 6.6 & 0.9 \\
59896.068647057 & -11.8 & 1.0 \\
59906.06373527 & 6.0 & 0.9 \\
\hline
\end{tabular}
\end{table}

\begin{figure}[h]
\begin{center}
\includegraphics[width=\linewidth]{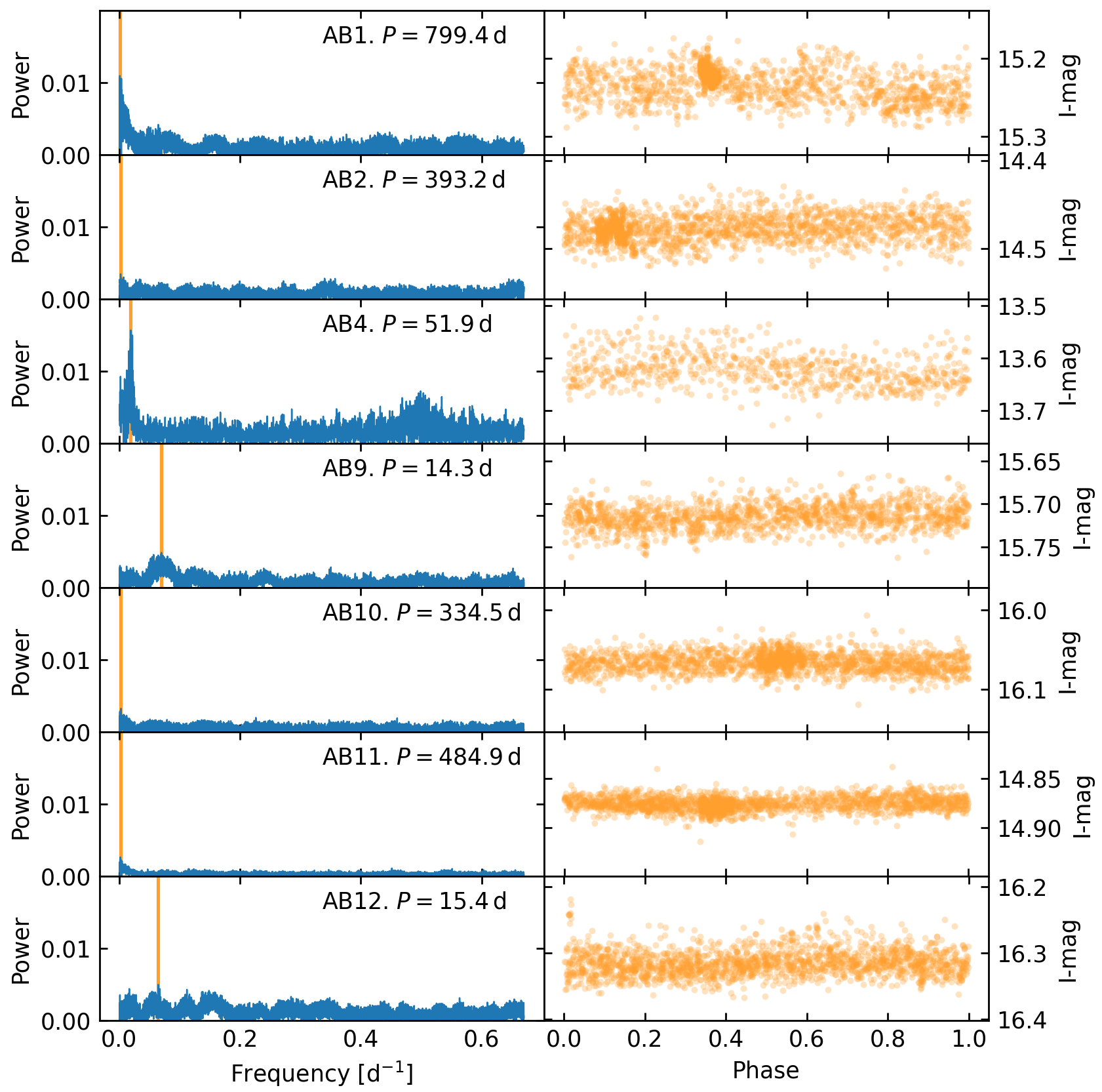} 
 \caption{ \textit{Left panels:} Lomb-Scargle periodograms of our target stars, based on OGLE data. The orange line shows the frequency at which the highest power is achieved, and the corresponding period is written in each panel. \textit{Right panels:} phase-folded light curves of I-band magnitudes, adopting the periods mentioned in the left panels.}
 \label{fig:periodograms}
\end{center}
\end{figure}

\begin{table}[t]
\caption{\label{tab:ogle}
Overview of the OGLE-III and OGLE-IV time-series photometry data that was analyzed in this work, showing the dates of the first and last observation, the number of observations ($N_\mathrm{obs}$), and the maximum SNR (SNR$_\mathrm{max}$) found for each source. No OGLE-IV data is available for SMC\,AB4.}
\small
\centering
\begin{tabular}{lllll}
\hline
\hline
Object & First observation & Last observation & $N_\mathrm{obs}$ & SNR$_\mathrm{max}$  \\
\hline
SMC\,AB1 & 07-2001 & 11-2022 & 1503 & 2.41\\
SMC\,AB2 & 06-2001 & 11-2022 & 1796 & 2.28\\
SMC\,AB4 & 06-2001 & 01-2009 & 718 & 2.47\\
SMC\,AB9 & 06-2001 & 11-2022 & 1441 & 2.40\\
SMC\,AB10 & 06-2001 & 11-2022 & 1785 & 2.48\\
SMC\,AB11 & 06-2001 & 11-2022 & 1889 & 2.89\\
SMC\,AB12 & 07-2001 & 11-2022 & 1764 & 2.26\\

\hline
\end{tabular}
\end{table}

\clearpage

\section{Detection probabilities \label{sec:appa}}

\begin{figure}[ht]
\begin{center}
\includegraphics[width=0.9\linewidth]{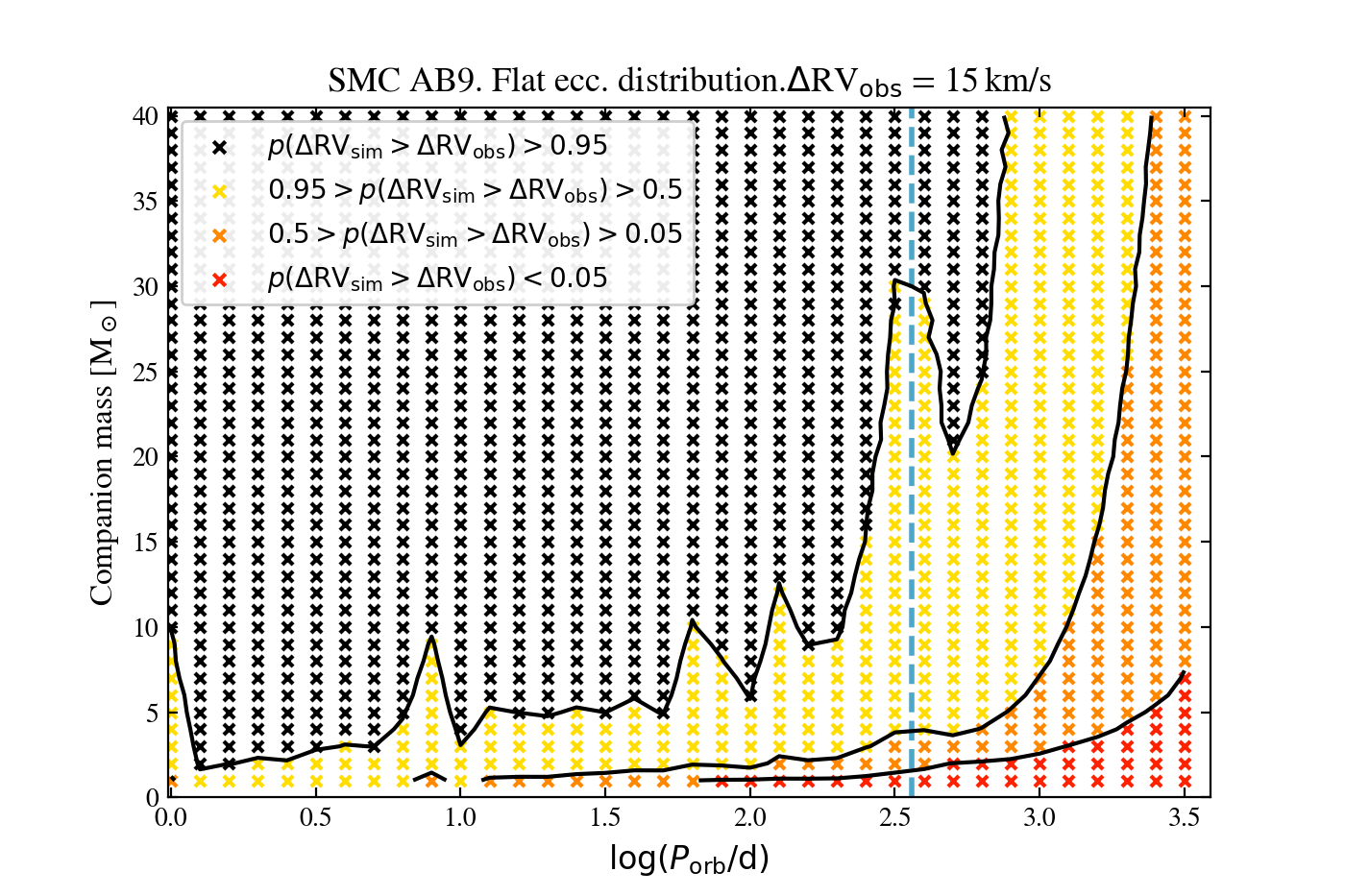} 
 \caption{ As Fig.\,\ref{fig:pbinary_ab9}, but assuming a flat eccentricity distribution instead of circular orbits.
 }
 \label{fig:pbinary_ab9_flatecc}
\end{center}
\end{figure}

\begin{figure}[ht]
\begin{center}
\includegraphics[width=0.9\linewidth]{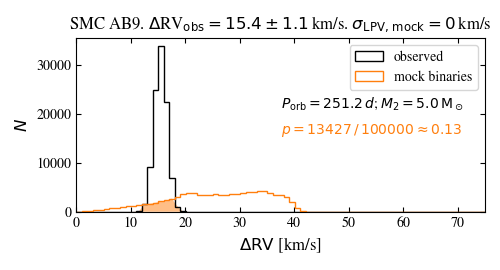} 
 \caption{ Histograms of simulated $\Delta$RV values for SMC\,AB9, where we show the scattered observations and a mock binary with $P_\mathrm{orb} = 501.2$\,d and a 5\,M$_\odot$ companion. }
 \label{fig:prob_ab9_mockbin}
\end{center}
\end{figure}

\begin{figure}[ht]
\begin{center}
\includegraphics[width=0.9\linewidth]{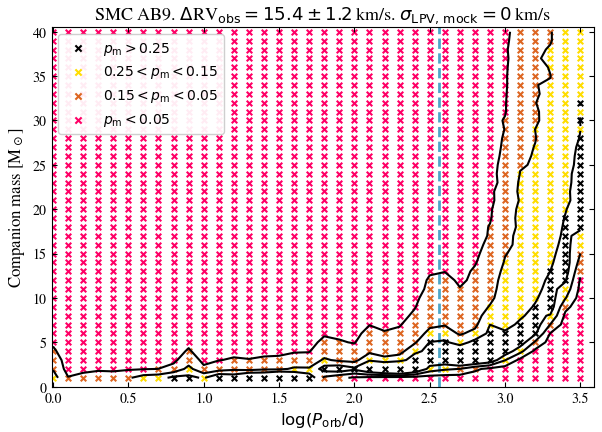} 
 \caption{ Same as Fig.\,\ref{fig:pbinary_ab9}, but instead of calculating the probability that $\Delta$RV$_\mathrm{sim}$ exceeds the threshold value $\Delta$RV$_\mathrm{obs}$, we calculate the probability $p_\mathrm{m}$ that $\Delta$RV$_\mathrm{sim}$ matches $\Delta$RV$_\mathrm{obs}$ (see text for details). }
 \label{fig:rvmatch_ab9}
\end{center}
\end{figure}

\subsection{More detection probability figures \label{sec:pbinary_figures}}

Figure\,\ref{fig:pbinary_ab9_flatecc} is the same figure as Fig.\,\ref{fig:pbinary_ab9}, but taking into account eccentricity. Specifically, eccentricities were drawn from a flat distribution between $e=0$ and $e=0.9$. The argument of periastron is given a random orientation for each mock binary.

We do another experiment where we try to match the observed $\Delta$RV value of SMC AB\,9 with binary motion, rather than using a $\Delta$RV cutoff. This is illustrated in Fig.\,\ref{fig:prob_ab9_mockbin}.
We first estimate the error on the observed $\Delta$RV by simulating our observing campaign of SMC\,AB9 10$^5$times. We give each RV measurement a random Gaussian offset based on its $\sigma_\mathrm{CCF}$, which causes a modest spread around the originally measured $\Delta$RV value (black line in Fig.\,\ref{fig:prob_ab9_mockbin}). The observed $\Delta$RV error is then inferred from fitting a Gaussian to the distribution of $\Delta$RV values. 
We then compare this with $\Delta$RV values from a mock binary with a certain orbital period and companion mass (e.g., $P_\mathrm{orb} = 251.2$\,d and a 5\,M$_\odot$ in Fig.\,\ref{fig:prob_ab9_mockbin}), for which we assume that the only uncertainty affecting $\Delta$RV is given by $\sigma_\mathrm{CCF}$.
Finally, we define $p_\mathrm{m}$ as the fractional overlap of the black and orange $\Delta$RV histograms (shaded orange region in Fig.\,\ref{fig:prob_ab9_mockbin}). Figure\,\ref{fig:rvmatch_ab9} shows the $p_\mathrm{m}$ values that we obtain over the whole binary parameter space in the range $0 \leq \log ( P_\mathrm{orb} / d ) \leq 3.5$ and $1 \leq M_2/M_\odot \leq 40$.

\subsection{Integrated detection probabilities \label{sec:pint}}

We can calculate the weighted average detection probability $\hat{p}$ by summing over all the probabilities $p_\mathrm{ij}$ we calculated (shown in Fig.\,\ref{fig:pbinary_ab9} and similar figures in Appendix\,\ref{sec:pbinary_figures}), weighted by the individual weights $w_\mathrm{P,i}$ (based on the orbital period) and $ w_\mathrm{M,j}$ (based the companion mass) and normalized by the sum of the individual weights:

\begin{equation}
    \hat{p} = \frac{\Sigma_{i,j=1,1}^{36, 40} \, p_\mathrm{i,j} w_\mathrm{P,i} w_\mathrm{M,j}}{ \Sigma_{i,j=1,1}^{36, 40} \, w_\mathrm{P,i} w_\mathrm{M,j} }.
\end{equation}
We sum from $i=1$ to $i=36$ because we explore 36 values for $\log ( P_\mathrm{orb} / d )$ in Sect.\,\ref{sec:possible_presence_of_companions}, and we sum over 40 values of $j$ because we explore 40 values for the companion mass. For flat probability distributions, individual weights $w_\mathrm{P,i}$ and $ w_\mathrm{M,j}$ have a value of unity if the orbital period and companion mass are within the range we consider (else they are zero). In two cases ($\hat{p}_\mathrm{L20}$ and $\hat{p}_\mathrm{R19}$), we use non-flat probability distributions of $\log ( P_\mathrm{orb} / d )$ and $M_\mathrm{2}$, which are based on previous binary populations synthesis works \citep{Langer20, Renzo19}. Below, we describe how we obtain these probability distributions.

To calculate $\hat{p}_\mathrm{L20}$, we used the orbital period and mass ratio distributions obtained from the population synthesis calculations of \cite{Langer20} for O star-black hole binaries in the LMC (cf., their figs.\,4 and\,6). The mass transfer in most of the underlying detailed binary models was highly inefficient. Since black hole formation kicks were neglected in this work, the predicted parameters of the O star-black hole binaries are expected to be similar the those of their O star-WR binaries progenitors.  

We also investigate the theoretical orbital period and mass ratio distribution based on models with conservative mass transfer \citep{Renzo19}, which produce wider binaries (more difficult to detect) with more massive companions (easier to detect).
To calculate $\hat{p}_\mathrm{R19}$, we used Engauge Digitizer to tabulate the semi-major axis distribution of $M > 15$\,M$_\odot$ accretor stars in fig.\,D.2 of \cite{Renzo19}. We converted this to an orbital period distribution assuming typical masses of  $M_\mathrm{WR} = 20$\,M$_\odot$ and  $M_\mathrm{comp} = 35$\,M$_\odot$. This lead to a period distribution with peaks around $\log (P_\mathrm{orb} / d) = 1$ and $\log (P_\mathrm{orb} / d) = 2.8$, going towards zero for $\log (P_\mathrm{orb} / d) < 0.2$ and $\log (P_\mathrm{orb} / d) > 3.5$. The companion mass distribution was estimated using the following assumptions: i) upon interaction, a 40\,M$_\odot$ star gets stripped to a 20\,M$_\odot$ WR star; ii) the minimum value of $q_\mathrm{crit} = 0.4$ for stable mass transfer adopted by \cite{Renzo19}; fully conservative mass transfer, i.e., the 20\,M$_\odot$ lost by the WR star end up on the accretor star; iv) a flat initial mass ratio distribution. This leads to a flat companion mass distribution between 36\,M$_\odot$ and 60\,M$_\odot$. To take a slightly conservative approach to estimate $\hat{p}_\mathrm{R19}$ and to bring this in line with the parameter space explored before, we instead adopt a flat companion mass distribution between 30\,M$_\odot$ and 40\,M$_\odot$.

We find that $\hat{p}_\mathrm{L20}$ and $\hat{p}_\mathrm{R19}$ are even higher ($\sim$98\%) than $\hat{p}$ ($\sim$90\%). The reason is that the binary models do not predict very low mass companions, and the binary models of \cite{Langer20} avoid orbital periods above $\log (P_\mathrm{orb} / d) = 3$. 

For both $\hat{p}_\mathrm{L20}$ and $\hat{p}_\mathrm{R19}$, we repeat the exercise but we take the systems with $P_\mathrm{orb} < 30$\,d out of the distribution, with the motivation that all of these might have been detected already. This shifts the probability distributions of the remaining binaries towards higher periods and hence the values for $\hat{p}_\mathrm{L20}$ and $\hat{p}_\mathrm{R19}$ become smaller. However, we still find average  detection probabilities in excess of 95\%.

\begin{table*}[t]
\caption{\label{tab:integrated_pdetect}
Probabilities that binaries would have caused a larger $\Delta$RV than observed. We integrated over different parts of the binary parameter space; if no extra restriction is mentioned, we consider the range $1 \leq M_\mathrm{2} / \mathrm{M}_\odot \leq 40$ and $0 \leq \log ( P_\mathrm{orb} / \mathrm{d} )\leq 3.5$. To calculate $\hat{p}_\mathrm{L20}$ and $\hat{p}_\mathrm{R19}$, we used orbital period and and mass ratio distributions based on \cite{Langer20} and \cite{Renzo19}, respectively (see text for details). }
\small
\centering
\begin{tabular}{llllllllll}
\hline \hline
& $\hat{p}$ & $\hat{p}_\mathrm{M5+}$ & $\hat{p}_\mathrm{P1yr-}$ & $\hat{p}_\mathrm{M5+P1yr-}$ & $\hat{p}_\mathrm{P30d+}$ &  $\hat{p}_\mathrm{L20}$ & $\hat{p}_\mathrm{L20,P30d+}$ & $\hat{p}_\mathrm{R19}$ & $\hat{p}_\mathrm{R19,P30d+}$ \\
 \hline
$M_\mathrm{2}$ restr. & & $M_\mathrm{2} >  5$  & & $M_\mathrm{2} >  5$ & & & & & \\ 
1$P_\mathrm{orb}$ restr. & & & $P_\mathrm{orb} <  1$\,yr & $P_\mathrm{orb} <  1$\,yr & $P_\mathrm{orb} > 30$\,d & & $P_\mathrm{orb} > 30$\,d & & $P_\mathrm{orb} > 30$\,d\\
\hline
SMC\,AB1 & 0.866 & 0.916 & 0.952 & 0.989 & 0.785 & 0.969 & 0.963 & 0.970 & 0.949\\
SMC\,AB2 & 0.927 & 0.963 & 0.978 & 0.996 & 0.880 & 0.990 & 0.988 & 0.988 & 0.980\\
SMC\,AB4 & 0.908 & 0.950 & 0.968 & 0.994 & 0.850 & 0.985 & 0.982 & 0.983 & 0.972\\
SMC\,AB9 & 0.887 & 0.916 & 0.967 & 0.993 & 0.798 & 0.985 & 0.982 & 0.965 & 0.942\\
SMC\,AB10& 0.956 & 0.974 & 0.995 & 0.999 & 0.924 & 0.998 & 0.997 & 0.992 & 0.986\\
SMC\,AB11& 0.892 & 0.929 & 0.971 & 0.994 & 0.822 & 0.983 & 0.980 & 0.973 & 0.955\\
SMC\,AB12& 0.852 & 0.896 & 0.956 & 0.988 & 0.760 & 0.972 & 0.966 & 0.954 & 0.924\\
\textbf{Average}& \textbf{0.901} & \textbf{0.935} & \textbf{0.970} & \textbf{0.993} & \textbf{0.831} & \textbf{0.983} & \textbf{0.980} & \textbf{0.975} & \textbf{0.958}\\
\hline
\end{tabular}
\end{table*}

\begin{table*}[t]
\caption{\label{tab:integrated_pdetect_ecc}
Same as Table\,\ref{tab:integrated_pdetect}, but considering \textbf{eccentric orbits} instead of only circular orbits. The eccentricity is drawn from a flat distribution between 0 and 0.9. }
\small
\centering
\begin{tabular}{llllllllll}
\hline \hline
& $\hat{p}$ & $\hat{p}_\mathrm{M5+}$ & $\hat{p}_\mathrm{P1yr-}$ & $\hat{p}_\mathrm{M5+P1yr-}$ & $\hat{p}_\mathrm{P30d+}$ &  $\hat{p}_\mathrm{L20}$ & $\hat{p}_\mathrm{L20,P30d+}$ & $\hat{p}_\mathrm{R19}$ & $\hat{p}_\mathrm{R19,P30d+}$ \\
 \hline
$M_\mathrm{2}$ restr. & & $M_\mathrm{2} >  5$  & & $M_\mathrm{2} >  5$ & & & & & \\ 
1$P_\mathrm{orb}$ restr. & & & $P_\mathrm{orb} <  1$\,yr & $P_\mathrm{orb} <  1$\,yr & $P_\mathrm{orb} > 30$\,d & & $P_\mathrm{orb} > 30$\,d & & $P_\mathrm{orb} > 30$\,d\\
\hline
SMC\,AB1 & 0.847 & 0.895 & 0.943 & 0.980 & 0.756 & 0.948 & 0.937 & 0.950 & 0.918\\
SMC\,AB2 & 0.911 & 0.946 & 0.970 & 0.991 & 0.853 & 0.978 & 0.973 & 0.977 & 0.963\\
SMC\,AB4 & 0.889 & 0.931 & 0.959 & 0.987 & 0.822 & 0.969 & 0.963 & 0.970 & 0.951\\
SMC\,AB9 & 0.856 & 0.895 & 0.955 & 0.984 & 0.769 & 0.966 & 0.960 & 0.946 & 0.913\\
SMC\,AB10& 0.941 & 0.960 & 0.991 & 0.998 & 0.901 & 0.994 & 0.993 & 0.984 & 0.973\\
SMC\,AB11& 0.870 & 0.908 & 0.959 & 0.985 & 0.790 & 0.964 & 0.956 & 0.955 & 0.926\\
SMC\,AB12& 0.831 & 0.874 & 0.941 & 0.976 & 0.729 & 0.945 & 0.933 & 0.933 & 0.890\\
\textbf{Average} & \textbf{0.878} & \textbf{0.916} & \textbf{0.960} & \textbf{0.986} & \textbf{0.803} & \textbf{0.966} & \textbf{0.968} & \textbf{0.959} & \textbf{ 0.933}\\
\hline
\end{tabular}
\end{table*}

\begin{table*}[t]
\caption{\label{tab:integrated_pdetect_f03} Same as Table\,\ref{tab:integrated_pdetect}, but for the observing campaign of \cite{Foellmi03} and, for SMC\,AB12, \cite{Foellmi04}. }
\small
\centering
\begin{tabular}{llllllllll}
\hline \hline
& $\hat{p}$ & $\hat{p}_\mathrm{M5+}$ & $\hat{p}_\mathrm{P1yr-}$ & $\hat{p}_\mathrm{M5+P1yr-}$ & $\hat{p}_\mathrm{P30d+}$ &  $\hat{p}_\mathrm{L20}$ & $\hat{p}_\mathrm{L20,P30d+}$ & $\hat{p}_\mathrm{R19}$ & $\hat{p}_\mathrm{R19,P30d+}$ \\
 \hline
$M_\mathrm{2}$ restr. & & $M_\mathrm{2} >  5$  & & $M_\mathrm{2} >  5$ & & & & & \\ 
1$P_\mathrm{orb}$ restr. & & & $P_\mathrm{orb} <  1$\,yr & $P_\mathrm{orb} <  1$\,yr & $P_\mathrm{orb} > 30$\,d & & $P_\mathrm{orb} > 30$\,d & & $P_\mathrm{orb} > 30$\,d\\
\hline
SMC\,AB1 & 0.479 & 0.537 & 0.660 & 0.740 & 0.229 & 0.401 & 0.282 & 0.634 & 0.413\\
SMC\,AB2 & 0.491 & 0.548 & 0.675 & 0.753 & 0.234 & 0.423 & 0.302 & 0.635 & 0.410\\
SMC\,AB4 & 0.526 & 0.584 & 0.716 & 0.794 & 0.278 & 0.491 & 0.380 & 0.679 & 0.481\\
SMC\,AB9 & 0.396 & 0.448 & 0.548 & 0.620 & 0.134 & 0.236 & 0.111 & 0.545 & 0.283\\
SMC\,AB10& 0.543 & 0.599 & 0.734 & 0.810 & 0.296 & 0.540 & 0.438 & 0.684 & 0.485\\
SMC\,AB11& 0.295 & 0.345 & 0.408 & 0.478 & 0.052 & 0.103 & 0.019 & 0.432 & 0.138\\
SMC\,AB12& 0.384 & 0.423 & 0.532 & 0.588 & 0.100 & 0.254 & 0.125 & 0.452 & 0.159\\
\textbf{Average}& \textbf{0.445} & \textbf{0.498} & \textbf{0.610} & \textbf{0.683} & \textbf{0.189} & \textbf{0.350} & \textbf{0.237} & \textbf{0.580} & \textbf{0.338} \\
\hline
\end{tabular}
\end{table*}

\begin{figure}[ht]
\begin{center}
\includegraphics[width=0.9\linewidth]{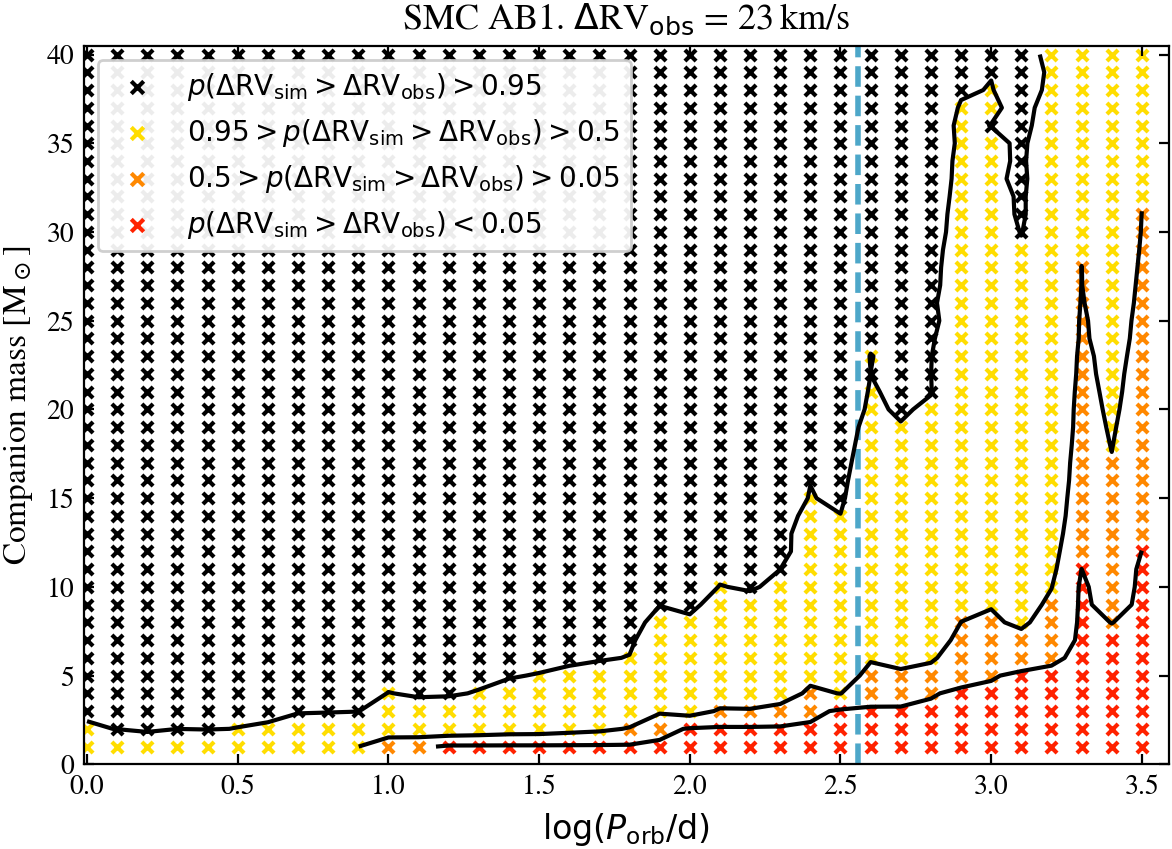} 
 \caption{ Same as Fig.\,\ref{fig:pbinary_ab9}, but for SMC\,AB1. }
 \label{fig:pdetect_ab1}
\end{center}
\end{figure}

\begin{figure}[ht]
\begin{center}
\includegraphics[width=0.9\linewidth]{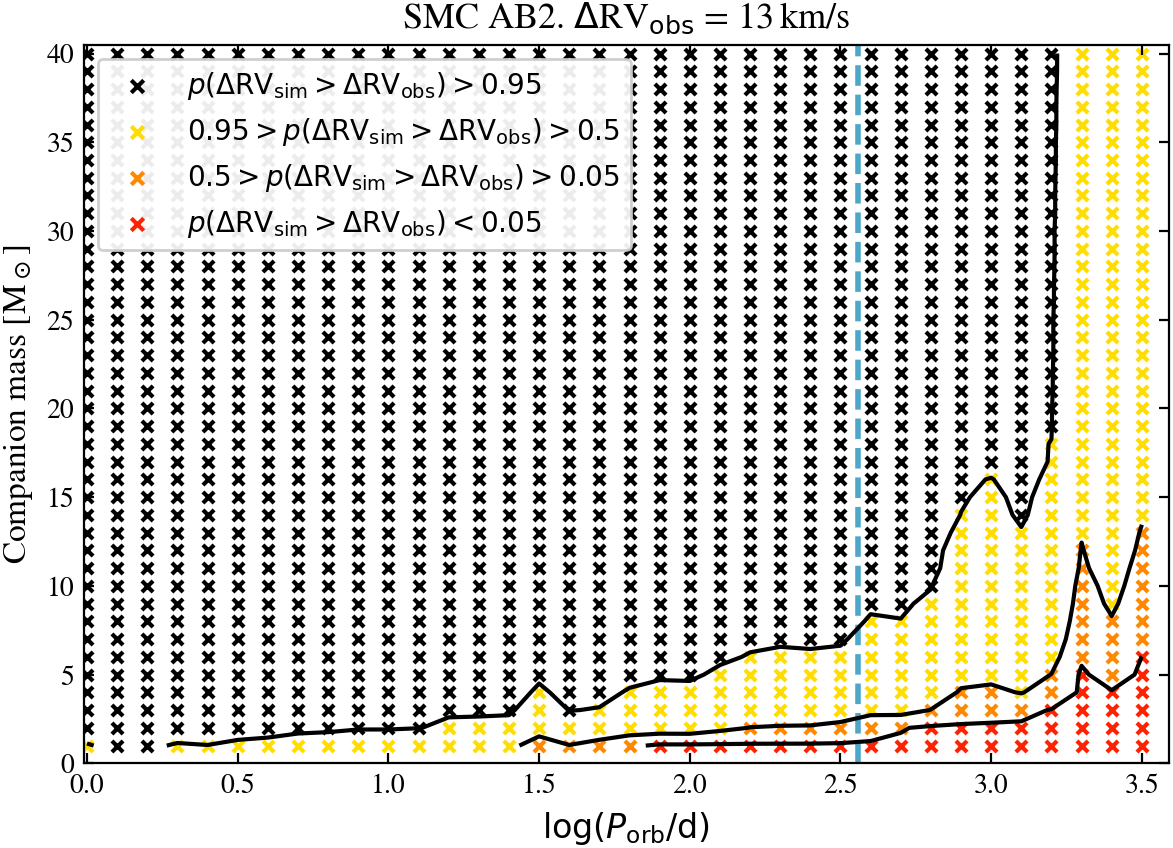} 
 \caption{ Same as Fig.\,\ref{fig:pbinary_ab9}, but for SMC\,AB2. }
\end{center}
\end{figure}

\begin{figure}[ht]
\begin{center}
\includegraphics[width=0.9\linewidth]{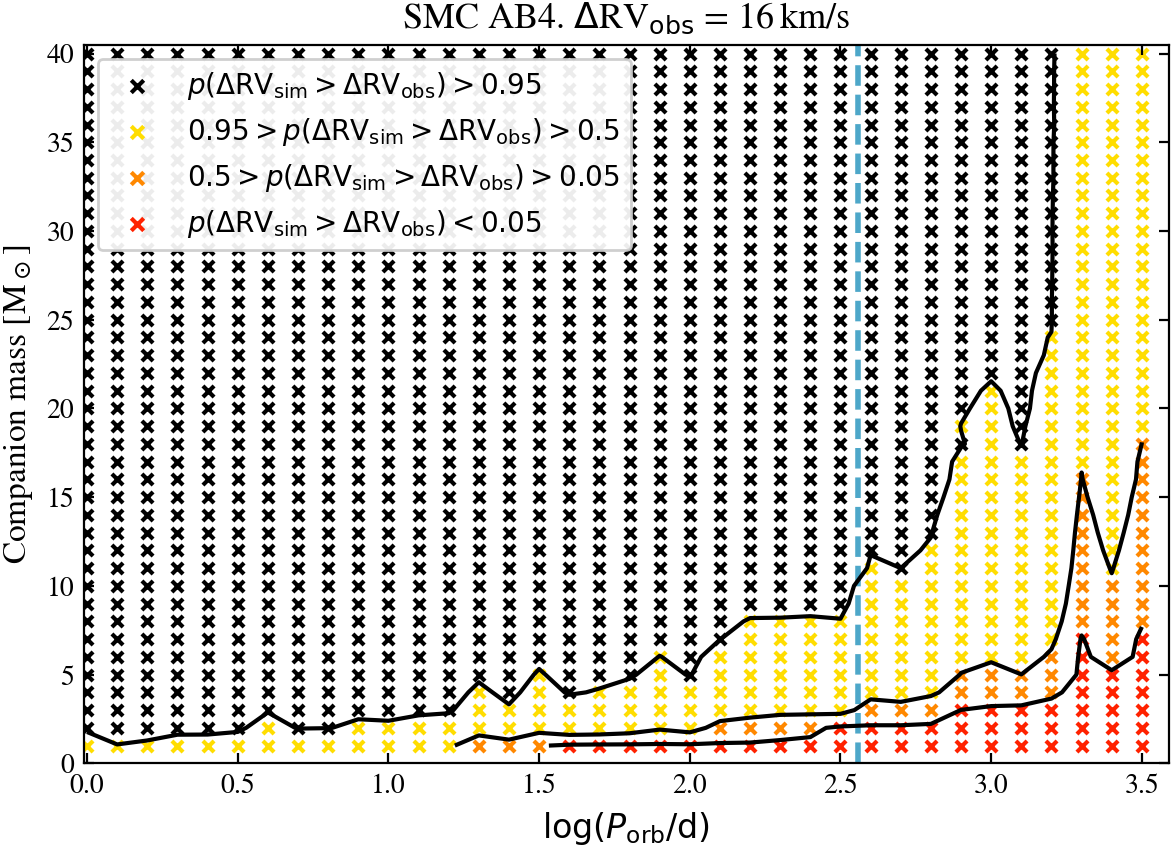} 
 \caption{ Same as Fig.\,\ref{fig:pbinary_ab9}, but for SMC\,AB4.
 }
\end{center}
\end{figure}

\begin{figure}[ht]
\begin{center}
\includegraphics[width=0.9\linewidth]{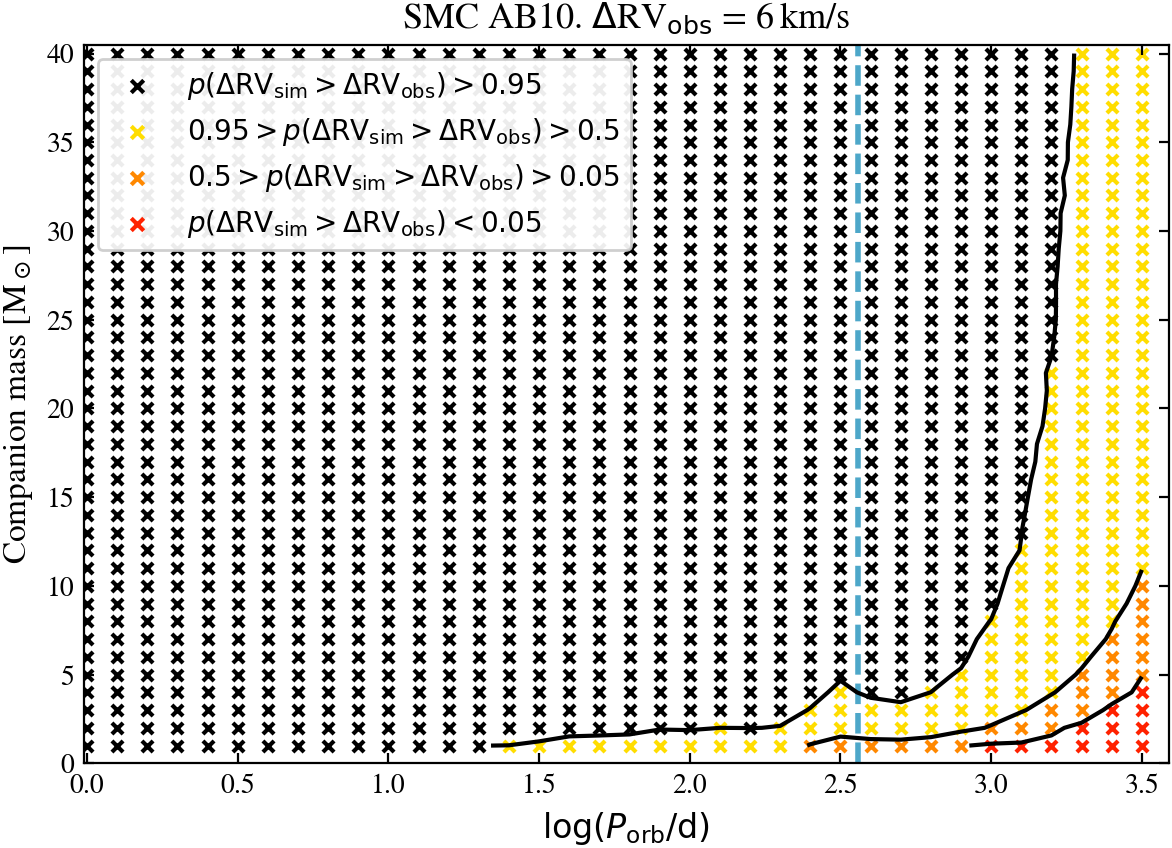} 
 \caption{ Same as Fig.\,\ref{fig:pbinary_ab9}, but for SMC\,AB10.
 }
\end{center}
\end{figure}

\begin{figure}[ht]
\begin{center}
\includegraphics[width=0.9\linewidth]{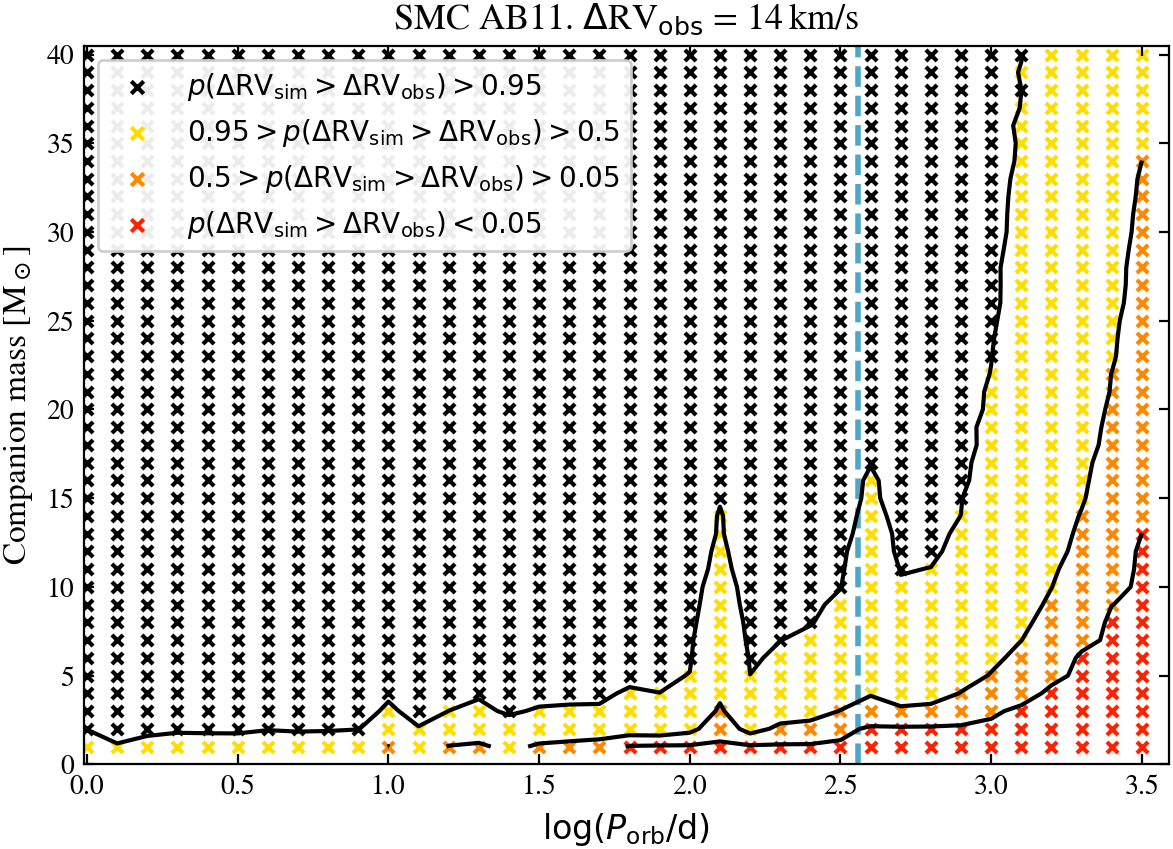} 
 \caption{ Same as Fig.\,\ref{fig:pbinary_ab9}, but for SMC\,AB11.
 }
\end{center}
\end{figure}

\begin{figure}[ht]
\begin{center}
\includegraphics[width=0.9\linewidth]{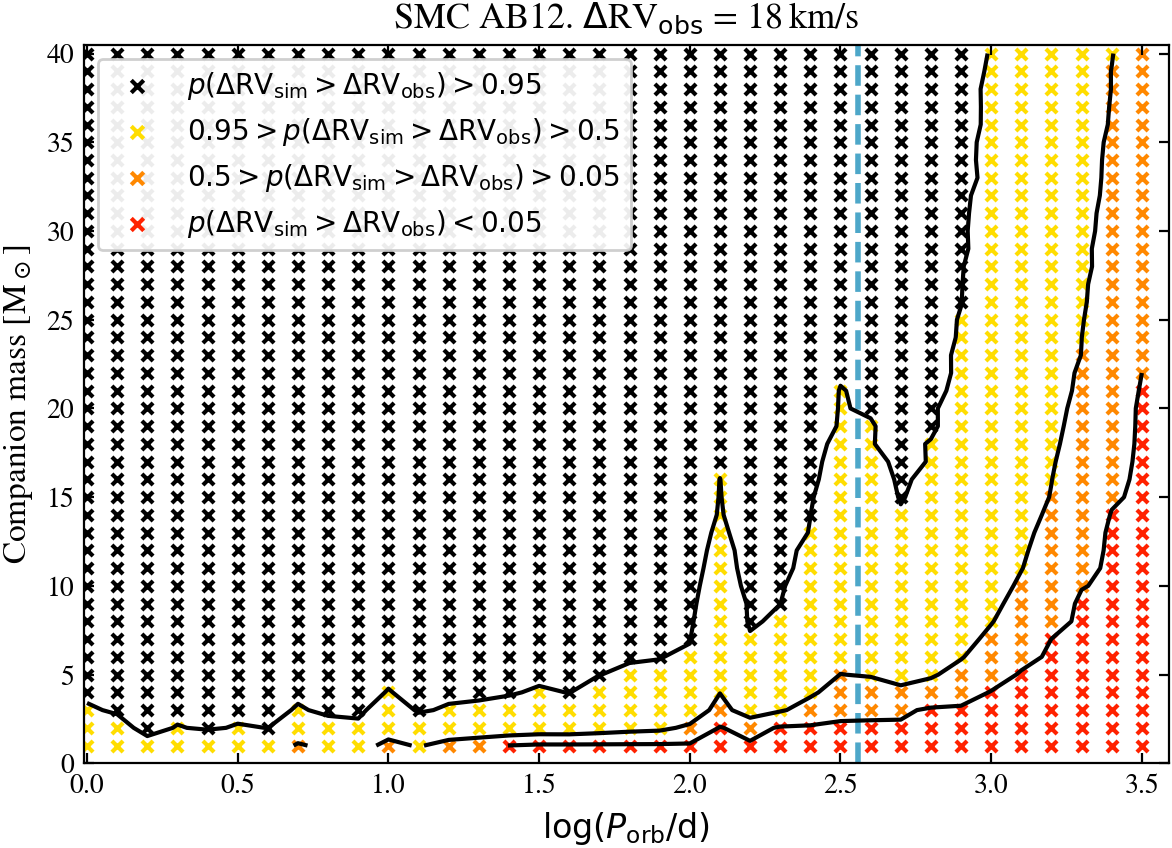} 
 \caption{ Same as Fig.\,\ref{fig:pbinary_ab9}, but for SMC\,AB12. }
  \label{fig:pdetect_ab12}
\end{center}
\end{figure}

\begin{figure}[ht]
\begin{center}
\includegraphics[width=0.9\linewidth]{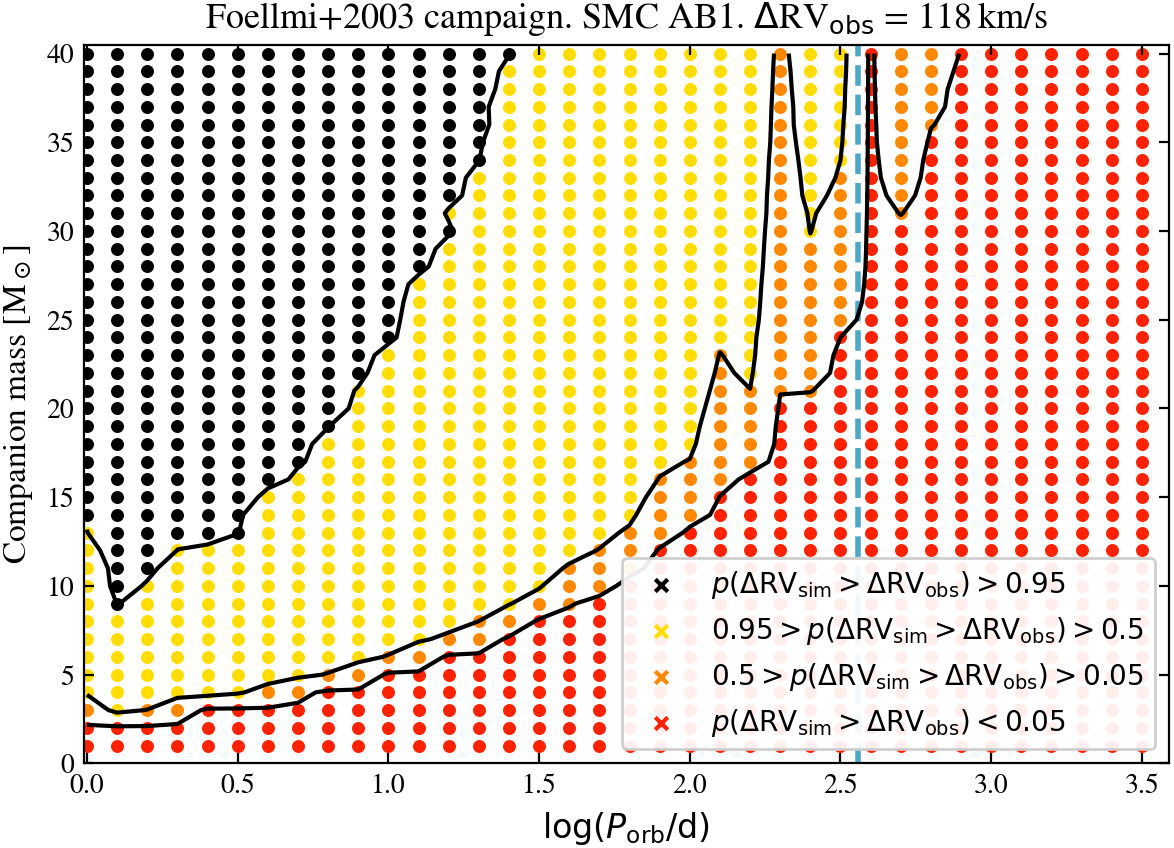} 
 \caption{ Probability that binary-induced motion would exceed the observed $\Delta \mathrm{RV}_\mathrm{obs}$ for measurements performed by F03 on SMC\,AB1.}
 \label{fig:pdetect_ab1_f03}
\end{center}
\end{figure}

\begin{figure}[ht]
\begin{center}
\includegraphics[width=0.9\linewidth]{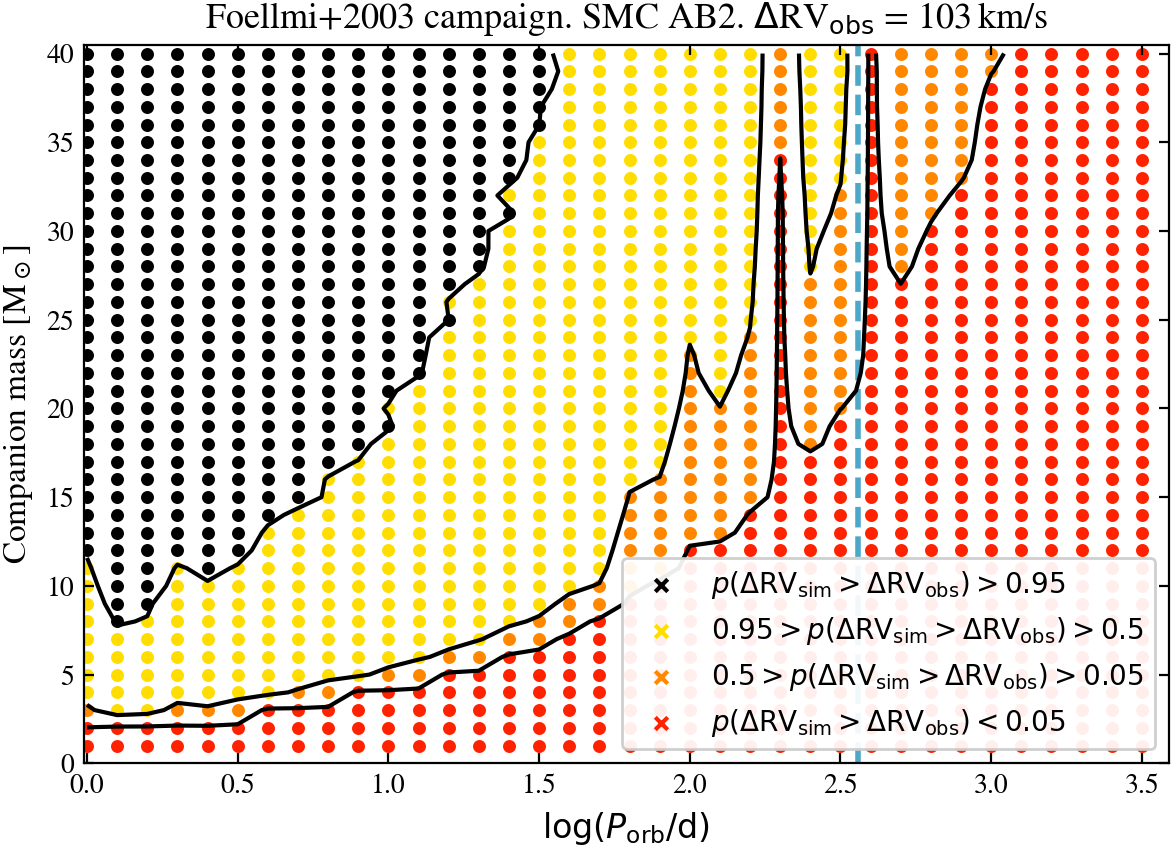} 
 \caption{ Same as Fig.\,\ref{fig:pdetect_ab1_f03}, but for SMC\,AB2.
 }
\end{center}
\end{figure}

\begin{figure}[ht]
\begin{center}
\includegraphics[width=0.9\linewidth]{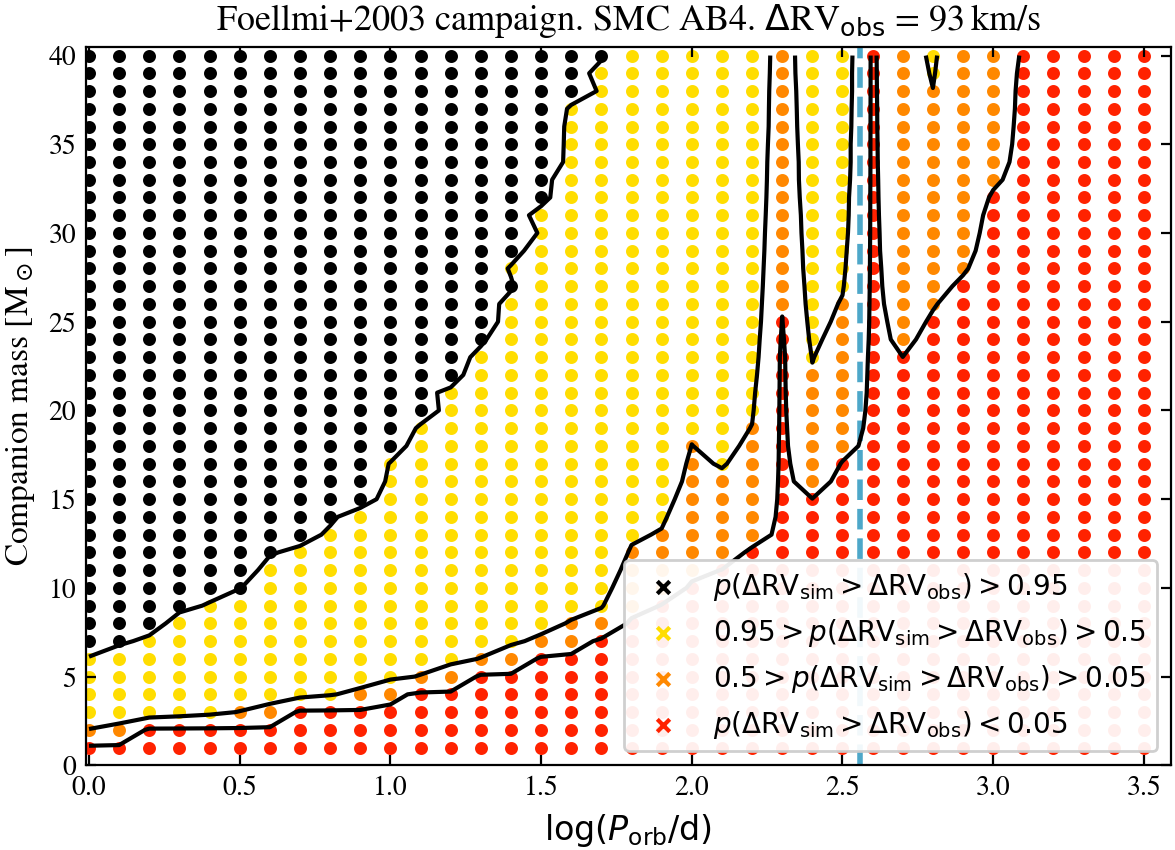} 
 \caption{ Same as Fig.\,\ref{fig:pdetect_ab1_f03}, but for SMC\,AB4. }
\end{center}
\end{figure}

\begin{figure}[ht]
\begin{center}
\includegraphics[width=0.9\linewidth]{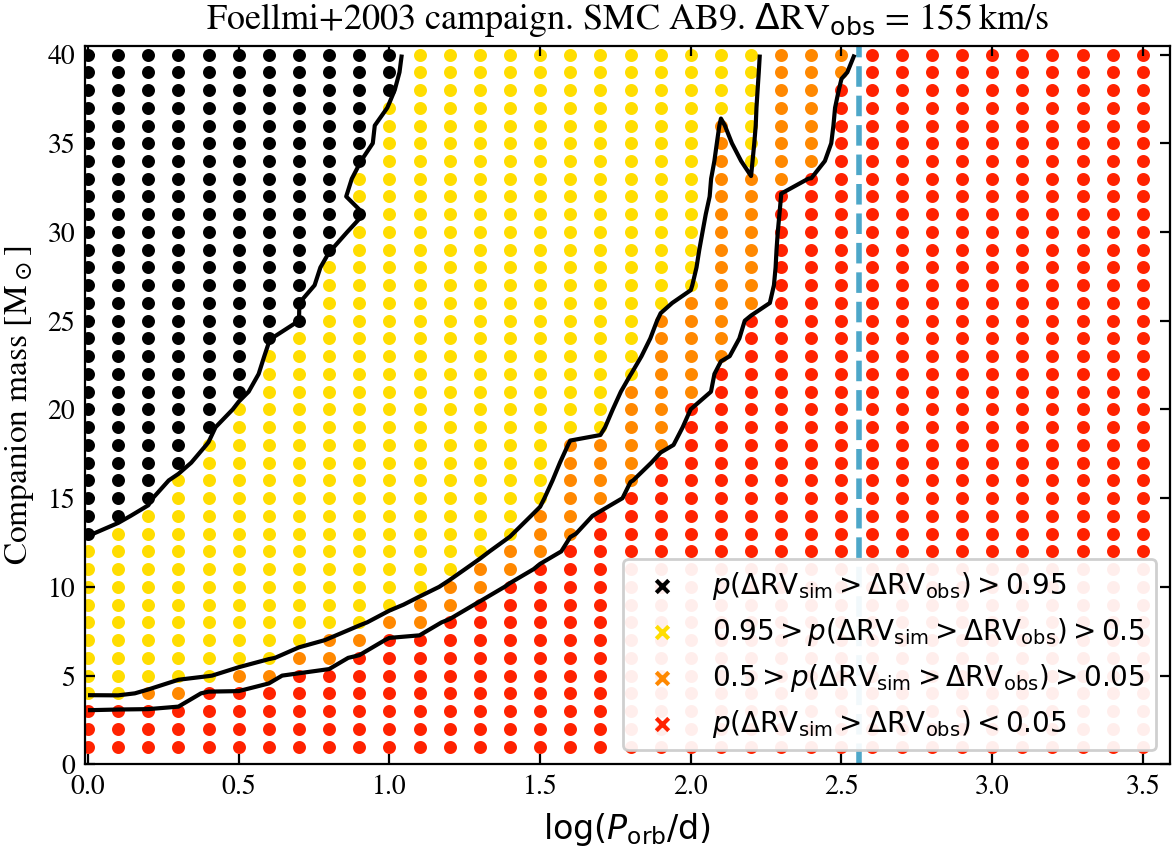} 
 \caption{ Same as Fig.\,\ref{fig:pdetect_ab1_f03}, but for SMC\,AB9. }
\end{center}
\end{figure}

\begin{figure}[ht]
\begin{center}
\includegraphics[width=0.9\linewidth]{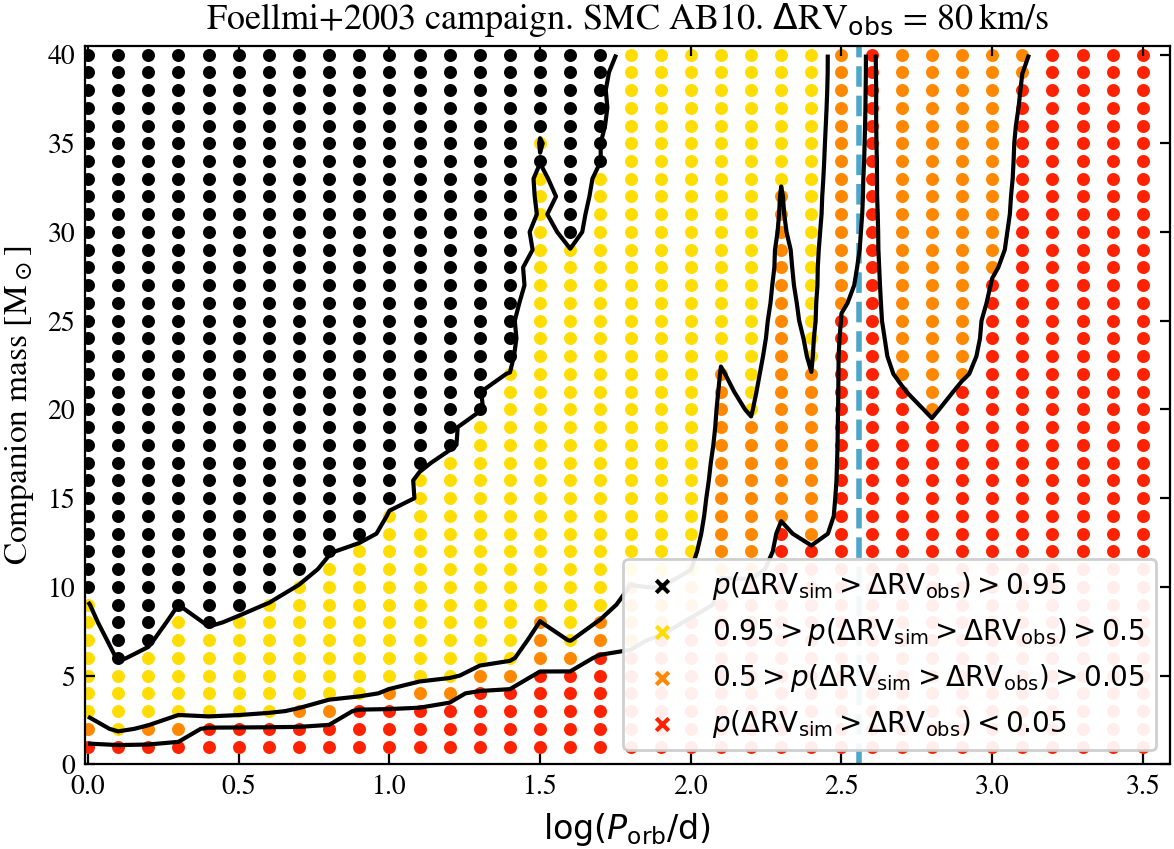} 
 \caption{ Same as Fig.\,\ref{fig:pdetect_ab1_f03}, but for SMC\,AB10. }
\end{center}
\end{figure}

\begin{figure}[ht]
\begin{center}
\includegraphics[width=0.9\linewidth]{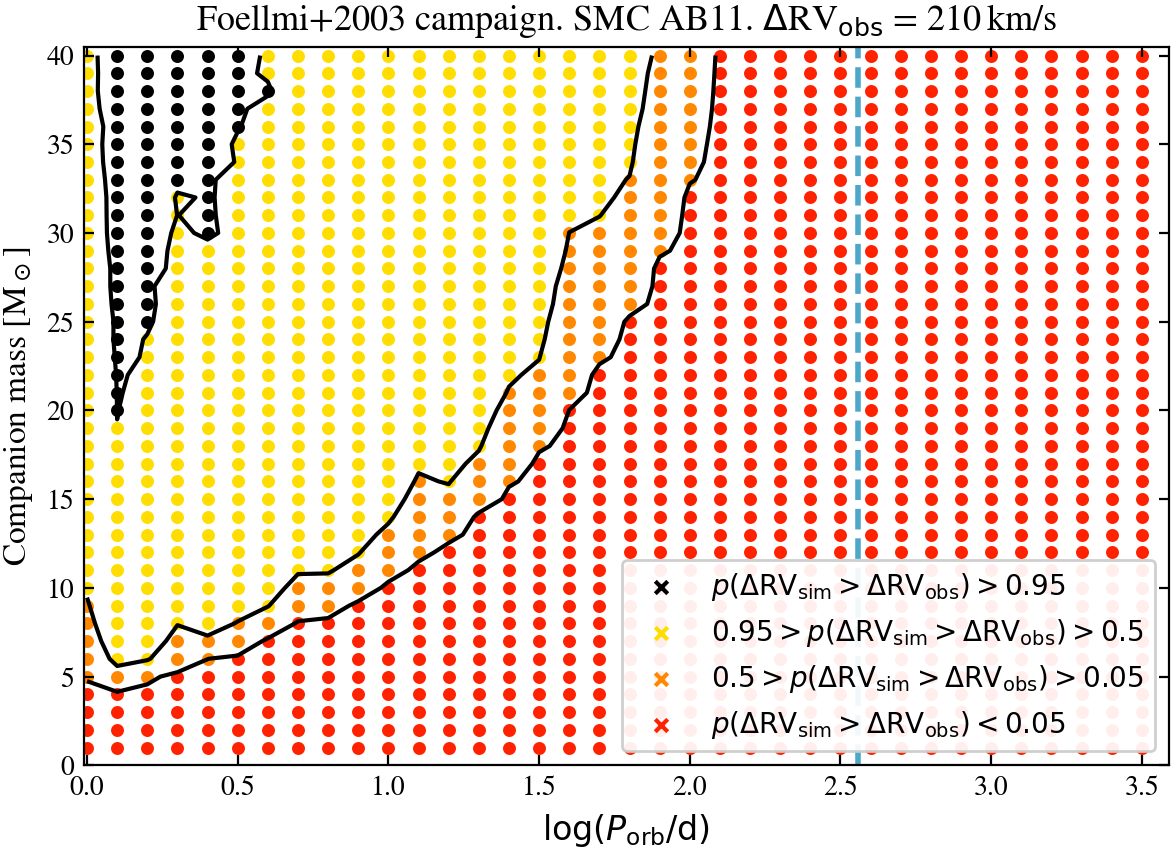} 
 \caption{ Same as Fig.\,\ref{fig:pdetect_ab1_f03}, but for SMC\,AB11. }
\end{center}
\end{figure}

\begin{figure}[ht]
\begin{center}
\includegraphics[width=0.9\linewidth]{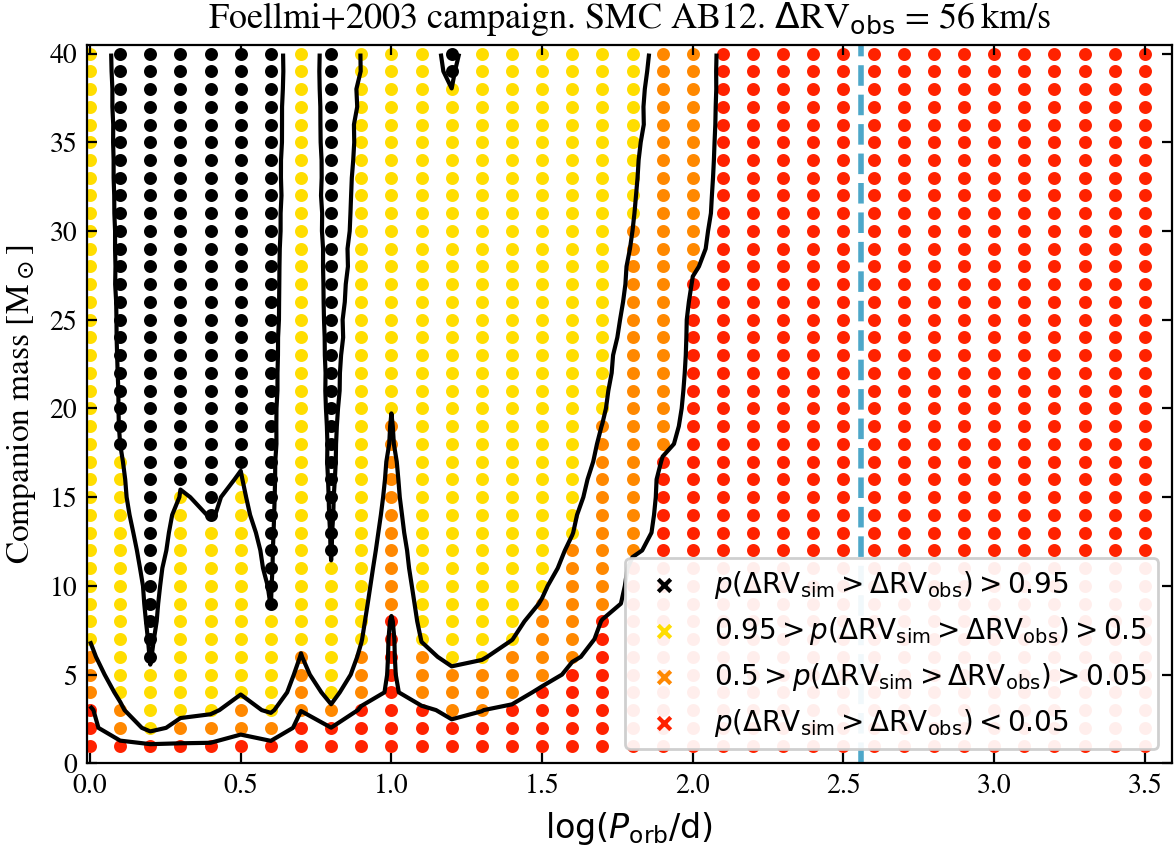} 
 \caption{ Same as Fig.\,\ref{fig:pdetect_ab1_f03}, but for SMC\,AB12. In this case, $\Delta \mathrm{RV}_\mathrm{obs}$ is based on measurements of \cite{Foellmi04}.}
  \label{fig:pdetect_ab12_f03}
\end{center}
\end{figure}

\end{document}